\documentclass[aps,prb,superscriptaddress,twocolumn,longbibliography]{revtex4-1}
\usepackage{bbm}
\usepackage{graphicx}
\usepackage{dcolumn}
\usepackage{bm}
\usepackage{subfigure}
\usepackage{amsmath}
\usepackage{feynmf}
\usepackage{hyperref}

\usepackage{float}
\usepackage{color}

\usepackage[titletoc]{appendix}

\usepackage{attachfile}

\newcommand{\bk}{\boldsymbol k}

\newcommand{\ba}{\boldsymbol a}
\newcommand{\bb}{\boldsymbol b}
\newcommand{\bn}{\boldsymbol n}
\newcommand{\bM}{\boldsymbol M}

\usepackage{times}

\begin{document}
\title{Sublattice-enriched tunability of bound states in second-order \\topological insulators
and superconductors}

\author{Di Zhu}
\affiliation{Guangdong Provincial Key Laboratory of Magnetoelectric Physics and Devices,
School of Physics, Sun Yat-Sen University, Guangzhou 510275, China}

\author{Majid Kheirkhah}
 \affiliation{Department of Physics, University of Alberta, Edmonton, Alberta T6G 2E1, Canada}
 \affiliation{Department of Physics, Simon Fraser University, Burnaby, British Columbia V5A 1S6, Canada}

\author{Zhongbo Yan}
\email{yanzhb5@mail.sysu.edu.cn}
\affiliation{Guangdong Provincial Key Laboratory of Magnetoelectric Physics and Devices,
School of Physics, Sun Yat-Sen University, Guangzhou 510275, China}

\date{\today}

\begin{abstract}
Bound states at sharp corners have been widely viewed as the hallmark of
two-dimensional second-order topological insulators and superconductors.
In this work, we show that the existence of sublattice degrees of freedom can
enrich the tunability of bound states on the boundary and hence
lift the constraint on their locations. We take the
Kane-Mele model with honeycomb-lattice structure to illustrate the underlying physics.
With the introduction of an in-plane exchange field  to the model, we find that
the boundary Dirac mass induced by the exchange field has a sensitive dependence on the boundary
sublattice termination. We find that the sensitive sublattice dependence can lead bound states to emerge at a specific
type of boundary defects named as
sublattice domain walls if the exchange field is of ferromagnetic nature,
even in the absence of any sharp corner on the boundary. Remarkably, this sensitive dependence of
the boundary Dirac mass on the boundary sublattice termination allows the positions of bound states to
be manipulated to any place on the boundary for an appropriately-designed sample. With a further introduction
of conventional s-wave superconductivity to the model, we find that, no matter whether the exchange field
is ferromagnetic, antiferromagnetic, or ferrimagnetic,  highly controllable Majorana zero modes
can be achieved at the sublattice domain walls.
Our work reshapes the understanding of boundary physics in second-order topological phases, and meanwhile
opens potential avenues to realize highly controllable bound states for potential applications.
\end{abstract}

\maketitle

\section{Introduction}

Since the discovery of two-dimensional (2D) topological insulators (TIs)~\cite{Kane2005a,Kane2005b,Bernevig2006a,Bernevig2006c,konig2007}, an
enduring and intensive exploration of topological phases in quantum materials
as well as various classical systems has been witnessed~\cite{hasan2010,qi2011,Ozawa2019review,Ma2019review}.
A hallmark of topological phases is the existence of gapless states
on the boundary enforced by the bulk topological invariant~\cite{Chiu2015RMP}. Conventionally,
the gapless states are known to be distributed on the boundary with the dimension lower than
the bulk by one. In other words, the gapless boundary states have codimension $d_{c}=1$.
Recently, it has been uncovered that there in fact exists a large class of topological phases
whose gapless boundary states have $d_{c}\geq2$~\cite{Sitte2012,Zhang2013surface,Benalcazar2017,Benalcazar2017prb,Schindler2018,Song2017,
Langbehn2017,Liu2017,ezawa2018higher,Geier2018,Khalaf2018,Yan2018hosc,Wang2018hosc,Wang2018weak,Yan2019hosca,Liu2019nh,Kudo2019,Chen2020hoti,
Huang2020FHOTI,Hu2020FHOTI,Zhang2021hotai}.
For distinction, now a topological phase is dubbed as an $n$th-order topological phase
if it only supports gapless boundary states with
$d_{c}=n$~\cite{Yan2019review,Schindler2020review,Xie2021}.

Different orders of topological phases have a hierarchy connection~\cite{Dumitru2019}.
In principle, an $n$th-order topological phase could be descended
from an $(n-1)$th-order topological phase by
appropriately lifting the protecting symmetry. A paradigmatic example is
the realization of a second-order TI by lifting the
time-reversal symmetry of a first-order TI~\cite{Sitte2012,Zhang2013surface,
Ren2020corner,Chen2020corner,Huang2022KM}. The physics
behind such a transition can be intuitively understood via the Jackiw-Rebbi theory based on
low-energy boundary Dirac-Hamiltonians~\cite{jackiw1976b,Yan2019review,Schindler2020review}.
That is, the breaking of time-reversal symmetry,
e.g., by applying a magnetic field, will introduce a boundary Dirac mass to gap out the
helical boundary (surface or edge) states, leading to a trivialization of
the first-order topological insulating phase~\cite{shen2013topological}.
Interestingly, the induced Dirac mass generally shows  a dependence on the orientation
of the boundary and may change sign across some direction~\cite{Khalaf2018}. When the Dirac masses on
two boundaries with different orientations have opposites signs, a Dirac-mass domain wall
harboring gapless states with $d_{c}=2$ will be formed at their intersection~\cite{jackiw1976b},
a corner in 2D~\cite{Ren2020corner,Chen2020corner,Huang2022KM,Zhu2018hosc,Zhuang2022}, or a hinge in 3D~\cite{Sitte2012,
Zhang2013surface,Zhang2019hinge,Majid2021vortex,Majid2021surface}.  Because of the generality of this domain-wall picture,
bound states positioned at corners in 2D systems~\cite{Peterson2018,Serra-Garcia2018,Imhof2018corner,Zhang2019sonic,Chen2019sotiPC,Fan2019corner,ElHassan2019}
and chiral or helical states propagating along hinges in 3D systems~\cite{Schindler2018bismuth,Gray2019helical,
Choi2020,Noguchi2021,Aggarwal2021,Shumiya2022} have been
widely taken as the defining boundary characteristic of second-order topology.

\begin{figure}[t]
\centering
\includegraphics[width=0.48\textwidth]{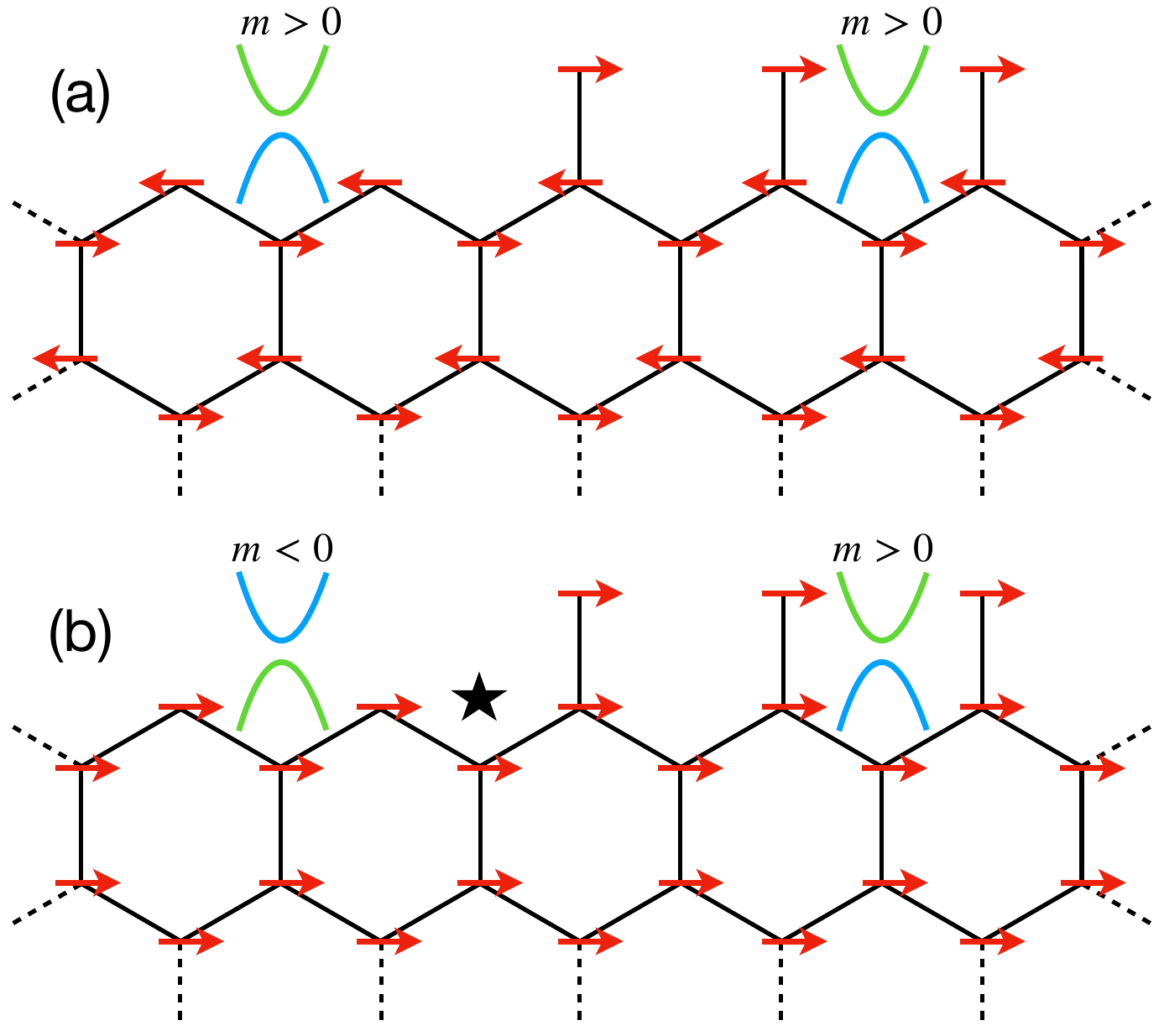}
\caption{(Color online) A schematic diagram of the sublattice domain walls and the dependence
of boundary Dirac mass ($m$) on the sublattice termination. The
upper boundary consists of two parts, with one part being a zigzag edge (top left hand side in each panel) and the other
being a beard edge (top right hand side in each panel), leading to the formation of sublattice domain walls at their intersections. The blue and green
parabolas represent massive Dirac energy spectra of the gapped edge states. Different color patterns
are just used to intuitively illustrate whether the signs of Dirac masses on the two sides of the sublattice domain wall are
the same [(a)] or opposite [(b)].
On each site, the red arrow
denotes the direction of the exchange field.
(a) When the exchange field is antiferromagnetic, the boundary Dirac masses on the
upper zigzag and beard edges have the same sign, accordingly, the sublattice domain wall is not a Dirac-mass
domain wall and hence does not harbor any bound states. (b) When the exchange field is ferromagnetic,
the boundary Dirac masses on the upper zigzag and beard edges have opposite signs, hence the
sublattice domain wall is a Dirac-mass domain wall supporting bound state illustrated by the black star.
}\label{sketch}
\end{figure}

When the sign of the boundary Dirac mass for a given system is
only sensitive to the orientation of the boundary, e.g., a higher-order
topological phase enforced by mirror symmetry~\cite{Langbehn2017,Geier2018}, it is true that the bound
states will be strongly bounded at sharp corners or hinges where
the orientation of the boundary has a dramatic change. However, if the boundary Dirac mass is also sensitive
to other factors on the boundary, then it is possible that the bound states are
not necessarily pinned at  sharp corners or hinges, but instead are allowed to
be positioned anywhere on the boundary. Obviously, the tunability of bound states
could make the observation of many interesting phenomena possible, such as
the creation of additional bound states or the annihilation of bound states.
Recently,
we did find that the sign of boundary Dirac mass in systems with sublattice degrees of
freedom can have a sensitive dependence on the boundary sublattice termination
in the context of second-order topological superconductors (TSCs)~\cite{Zhu2021sublattice,Majid2022kagome}.
Concretely, we found that when a 2D first-order TI with honeycomb~\cite{Zhu2021sublattice} or kagome
 lattice structure~\cite{Majid2022kagome} is placed
on an unconventional superconductor, e.g., a d-wave superconductor,  the Dirac mass
induced by superconductivity gapping out the helical edge states exhibits
a sensitive dependence on the type of terminating sublattices on the boundary.
This property allows the realization of Majorana Kramers pairs (a Majorana
Kramers pair corresponds to two Majorana zero modes (MZMs) related by time-reversal symmetry~\cite{Haim2019review})
at the so-called sublattice domain walls, a type of boundary defects
corresponding to the intersection of two edges with the same
orientation but with different sublattice terminations, see illustration in
Fig.\ref{sketch}. Remarkably,
the Majorana Kramers pairs with $d_{c}=2$ can emerge even without the existence of sharp
corners (e.g., a cylindrical geometry with one direction being periodic)~\cite{Zhu2021sublattice}, and
their positions can be manipulated by tuning the sublattice
terminations~\cite{Zhu2021sublattice,Majid2022kagome}, which may benefit the future
application of Majorana bound states in topological quantum computation~\cite{nayak2008review,Liu2014Majorana,gao2016symmetry,Schrade2018MKPs}.

It is known that when the time-reversal symmetry is broken, the helical boundary
states of a first-order TI would be gapped out~\cite{Yu2010QAHE,qi2011}. The time-reversal symmetry could be broken by
an exchange field, which itself could be induced by an emergent intrinsic magnetic order
or magnetic proximity effect from a substrate magnetic insulator.  As Dirac domain walls
can exist in both superconductors and insulators,
it is natural to expect that the Dirac mass induced by exchange field
may also have similar sensitive dependence on the sublattice termination,
and the realization of bound states at sublattice domain walls
may also occur in the context of second-order TIs.
In this work, we take the paradigmatic Kane-Mele model
with honeycomb lattice structure to demonstrate this
expectation~\cite{Kane2005a,Kane2005b}. The Kane-Mele
model is known to support first-order topological
insulating phase, and the honeycomb lattice contains only
two sublattice degrees of freedom (labeled as A and B for discussion).
Following previous works, we consider that the exchange field
lies in the lattice plane~\cite{Ren2020corner,Chen2020corner},
and for generality  we consider that the collinear magnetic moments on
the two types of sublattices satisfy $\bM_{A}=\gamma \bM_{B}$ with $-1\leq\gamma\leq1$.
Correspondingly,
$0\leq\gamma\leq1$ refers to a ferromagnetic order, $\gamma=-1$ refers to an antiferromagnetic order,
and $-1<\gamma<0$ refers to a ferrimagntic order. Based on the
low-energy edge theory~\cite{Yan2018hosc,Zhu2021sublattice}, we determine the boundary Dirac masses on
the two types of edges whose terminations contain only one type
of sublattices (commonly dubbed zigzag and beard edges~\cite{neto2009}).

Our main findings can be briefly summarized as follows.
First, we find that, for the zigzag and beard edges with the same orientation,
whether the values or signs of the Dirac masses on them
are the same or not depends on the value of $\gamma$.
Somewhat counterintuitively, we find that, for
an antiferromagnetic or ferrimagnetic exchange field, i.e., $-1\leq\gamma<0$,  the Dirac masses
on them  take the same sign, even though the directions of the exchange field are opposite
on the outermost terminating sublattices for these two kinds of
edges, as depicted in Fig.\ref{sketch}(a). On the contrary, for a ferromagnetic
exchange field, we find that
the Dirac masses on them take opposite signs,
even though the directions of the exchange field are the same
on the outermost terminating sublattices for these two kinds of
edges, as depicted in Fig.\ref{sketch}(b). Because of the sign difference in Dirac masses,
we find that the ferromagnetic exchange field can induce highly controllable bound states at the
sublattice domain walls corresponding to the intersection of
zigzag and beard edges. As an important consequence,
bound states can be achieved even in the absence of any sharp corner on the boundary.
As the boundary Dirac mass induced by exchange field and the effective chemical potential
on the boundary
turn out to have a sensitive dependence on the sublattice terminations,
we  show that these properties allows the realization of MZMs
 at the sublattice domain walls even one considers conventional s-wave superconductivity
which, in the absence of exchange field, will introduce a uniform boundary Dirac mass~\cite{Zhu2021sublattice}.
Our findings suggest that the ubiquitous sublattice degrees of
freedom in materials provide a knob to control and manipulate the
positions of bound states in second-order topological phases.

This paper is organized as follows. In Sec.\ref{sec2}, we introduce the Hamiltonian describing
a first-order TI subjected to an in-plane exchange field. In Sec.\ref{sec3},
we establish a theory distinct from the one developed
by Ren {\it et al.}~\cite{Ren2020corner} to understand the robustness of helical edge states
on the armchair edges, and show that the helical edge states on the armchair edges will be gapped
out once the exchange fields on the two sublattices are different, leading to the presence
of corner bound states in samples with geometries different from the one with diamond shape considered in Ref.~\cite{Ren2020corner}.
In Sec.\ref{sec4}, we derive the low-energy boundary Hamiltonians on the beard and zigzag
edges,  and show explicitly the dependence of boundary Dirac masses on the sublattice terminations.
The presence of bound states at the sublattice domain walls is also numerically demonstrated.
In Sec.\ref{sec5}, we further consider the introduction of s-wave superconductivity to the
system and show the presence of MZMs at the sublattice domain walls.
We discuss the results and conclude the paper in Sec.\ref{sec6}.
Some calculating details of the low-energy boundary Hamiltonians
are relegated to appendices.

\section{Kane-Mele model with an in-plane exchange field}
\label{sec2}

We start with the Hamiltonian~\cite{Kane2005a,Kane2005b}
\begin{eqnarray}
H&=&t\sum_{\langle ij\rangle,\alpha}c_{i,\alpha}^{\dag}c_{j,\alpha}+i\lambda_{\rm so}\sum_{\langle\langle ij\rangle\rangle,\alpha,\beta}\nu_{ij}c_{i,\alpha}^{\dag}(s_{z})_{\alpha\beta}c_{j,\beta}\nonumber\\
&&+\lambda_{\nu}\sum_{i,\alpha}\xi_{i}c_{i,\alpha}^{\dag}c_{i,\alpha}+\sum_{i,\alpha,\beta}c_{i,\alpha}^{\dag}(\bM_{i}\cdot \mathbf{s})_{\alpha\beta}c_{i,\beta},
\end{eqnarray}
where $c_{i,\alpha}^{\dag}$ $(c_{i,\alpha})$ refers to a fermion creation (annihilation)
operator at site $i$, the subscripts $\alpha$ and $\beta$ refer to spin indices, $t$ denotes the hopping energy, $\lambda_{\rm so}$ characterizes
the strength of intrinsic spin-orbit coupling, $\nu_{ij}=1(-1)$ for a clockwise (anticlockwise)
path along which the electrons hop from site $j$ to site $i$, $\lambda_{\nu}$ characterizes
the staggered sublattice potential ($\xi_{i}=\pm1$), and the last term describes the exchange field
induced by certain collinear magnetic order (the involving $g$-factor and $\hbar$ are made implicity for
notational simplicity). The notations $\langle ij\rangle$ and $\langle\langle ij\rangle\rangle$
mean that the sum is over nearest-neighbor sites and next-nearest-neighbor sites, respectively. As
in this paper we are interested in second-order topology, the collinear magnetic order will be assumed to lie
in the lattice plane, i.e., $\bM_{i\in A}=\bM=(M_{x},M_{y},0)$,
and for generality we consider $\bM_{i\in B}=\gamma\bM$ with $-1\leq\gamma\leq1$
to take all possible in-plane collinear magnetic orders into account.

By performing a Fourier transformation and choosing
the basis to be $\psi_{k}=(c_{k,A,\uparrow},c_{k,B,\uparrow},c_{k,A,\downarrow},c_{k,B,\downarrow})^{T}$,
the Hamiltonian in momentum space reads
\begin{eqnarray}
\mathcal{H}(\bk)&=&t\sum_{i=1}^{3}[\cos(\bk\cdot\ba_{i})s_{0}\sigma_{x}+\sin(\bk\cdot\ba_{i})s_{0}\sigma_{y}]\nonumber\\
&&+2\lambda_{\rm so}\sum_{i=1}^{3}\sin(\bk\cdot\bb_{i})s_{z}\sigma_{z}+\lambda_{\nu}s_{0}\sigma_{z}\nonumber\\
&&+\frac{1+\gamma}{2}(\bM\cdot \mathbf{s})\sigma_{0}+\frac{1-\gamma}{2}(\bM\cdot \mathbf{s})\sigma_{z},\label{normal}
\end{eqnarray}
where the Pauli matrices $s_{i}$ and $\sigma_{i}$, and the identity matrices
$s_{0}$ and $\sigma_{0}$,  act on
the spin ($\uparrow,\downarrow$) and sublattice (A,B)
degrees of freedom, respectively.
$\ba_{i}$ refers to
the nearest-neighbor vectors, with $\ba_{1}=a(0,1)$, $\ba_{2}=\frac{a}{2}(\sqrt{3},-1)$,
$\ba_{3}=\frac{a}{2}(-\sqrt{3},-1)$ (throughout the paper we set the lattice constant $a=1$ for notational
simplicity). The next-nearest-neighbor vectors
$\bb_{1}=\ba_{2}-\ba_{3}$, $\bb_{2}=\ba_{3}-\ba_{1}$ and $\bb_{3}=\ba_{1}-\ba_{2}$~\cite{Haldane1988}.
The last line in (\ref{normal}) means that a general collinear exchange field can be decomposed as the sum of a
uniform ferromagnetic exchange field and an antiferromagnetic one.
Without the two time-reversal-symmetry-breaking terms in the last line, the Hamiltonian describes a first-order TI when $|\lambda_{\nu}|<3\sqrt{3}|\lambda_{\rm so}|$~\cite{Kane2005b}.

\section{Helical edge states and corner bound states on the boundary}
\label{sec3}

Considering the Kane-Mele model with only the ferromagnetic term, Ren {\it et al.} showed
that the helical edge states would be gapped
out on the zigzag edges, but remain gapless on the armchair edges~\cite{Ren2020corner}, irrespective
of the direction of the in-plane ferromagnetic exchange field.
In order to avoid gapless edge states with codimension $d_{c}=1$
and only have bound states with $d_{c}=2$ on the boundary, Ren {\it et al.} suggested a diamond-shaped
nanoflake with only zigzag boundaries. For such a geometry, they showed that
helical edge states are gapped out on all edges, while  bound states show up
at half of the corners~\cite{Ren2020corner}. By a close look of the Fig.1 therein, one can notice that
these corners hosting bound states correspond to the intersections of  two adjacent zigzag
edges with different orientations as well as distinct sublattice terminations.

\begin{figure}[t]
\centering
\includegraphics[width=0.48\textwidth]{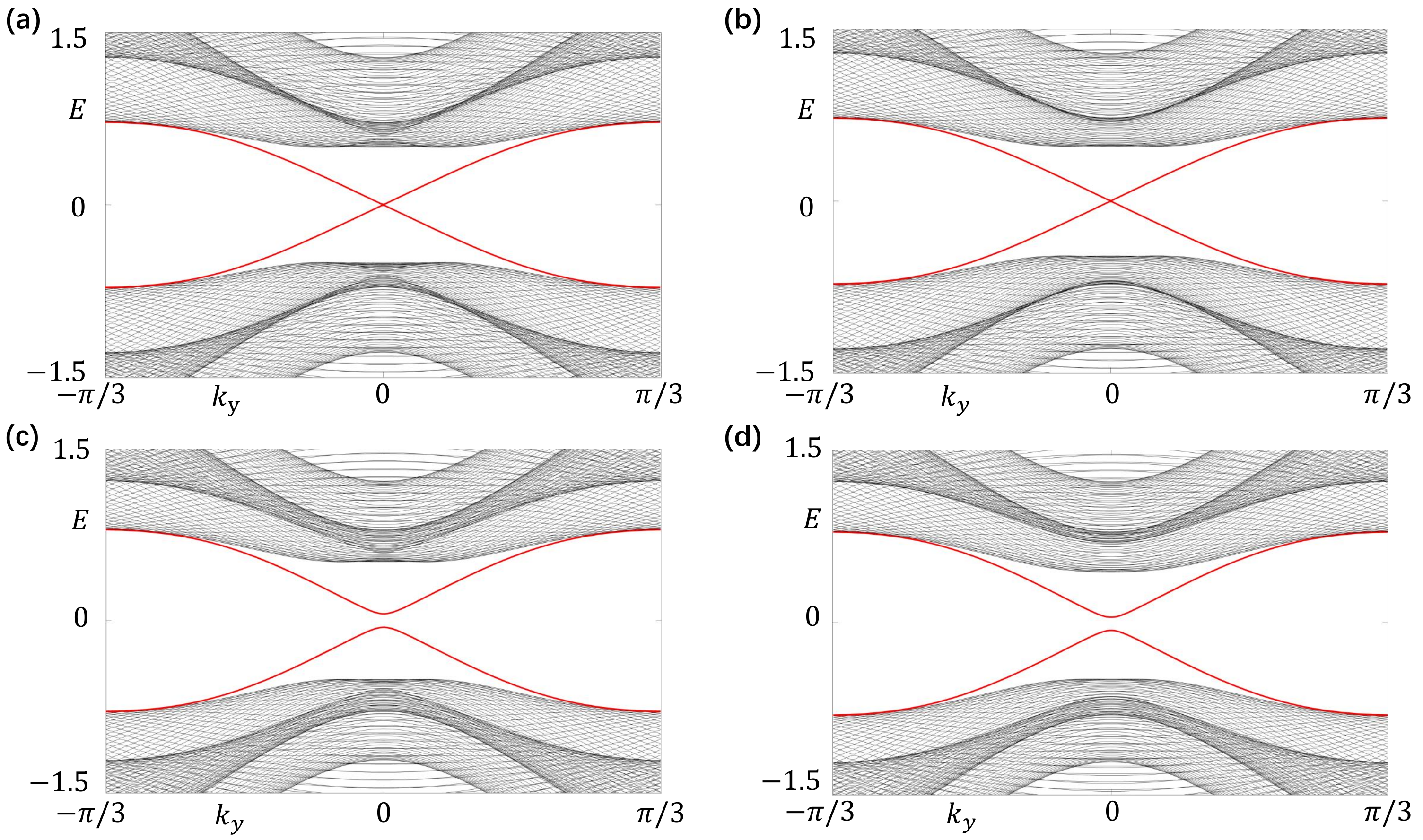}
\caption{(Color online) Energy spectra for a ribbon with armchair edges.
The ribbon has open boundary conditions in the $x$ direction and periodic boundary
conditions in the $y$ direction.
Chosen parameters are $t=1$, $\lambda_{\rm so}=0.1$,
$M_{x}=M_{y}=0.2$. (a)  $\lambda_{\nu}=0$, $\gamma=1$, (b) $\lambda_{\nu}=0.1$, $\gamma=1$,
(c) $\lambda_{\nu}=0$, $\gamma=0.5$, (d) $\lambda_{\nu}=0.1$, $\gamma=0.5$.  }\label{armchair}
\end{figure}

In Ref.~\cite{Ren2020corner}, the reason that the helical edge states are stable against
the in-plane exchange field (corresponding to $\gamma=1$) was attributed to the existence of an additional
mirror symmetry on the armchair edges, which provides a further protection. In more detail,
it is known that $k_{y}$ remains a good quantum number on the armchair edges if periodic boundary conditions
are imposed in the $y$ direction. In
Ref.~\cite{Ren2020corner}, the authors considered the limit without the staggered sublattice
potential, i.e., $\lambda_{\nu}=0$, and found that the reduced Hamiltonian
$\mathcal{H}(k_{x},k_{y}=0)$ has a mirror symmetry, with the symmetry operator taking  the general form
$\mathcal{M}_{n}=i(\hat{\bn}\cdot \mathbf{s})\sigma_{x}$,
where $\hat{\bn}$ denotes the direction of the magnetic moment. As $[\mathcal{H}(k_{x},0),\mathcal{M}_{n}]=0$,
$\mathcal{H}(k_{x},0)$ can be decomposed as a direct sum of two sectors for
any $k_{x}$ according to the two eigenvalues ($\pm i$)
of the mirror operator, and it turns out that the two sectors carry
opposite winding numbers, $\omega^{\pm i}=\pm1$. Although this explanation is valid
in the limit with $\lambda_{\nu}=0$, it is in fact not essential. To see this,
we would like to point out that the helical edge states in fact
remain gapless when $\lambda_{\nu}\neq0$, as long as
the exchange field has the same value on the two types of sublattices, i.e., $\gamma=1$, as shown in
Figs.\ref{armchair}(a)(b). However, once
$\lambda_{\nu}\neq0$, the aforementioned mirror symmetry is explicitly broken
by the staggered sublattice potential (note $[\lambda_{\nu}s_{0}\sigma_{z},\mathcal{M}_{n}]\neq0$).
Accordingly, the Hamiltonian can no longer be decomposed into two mirror sectors,
and the topological analysis based on the mirror-graded winding numbers in Ref.~\cite{Ren2020corner} breaks down.
Nevertheless, the robustness of the crossing at $k_{y}=0$ on the armchair edges
even when $\lambda_{\nu}\neq0$ suggests
the existence of a topological protection. Viewing $\mathcal{H}(k_{x},0)$ as
a one-dimensional Hamiltonian, the double degeneracy of the crossing on
one armchair edge suggests the existence of two zero-energy bound states
at each end of the one-dimensional system. As the spinful time-reversal
symmetry is explicitly broken by the exchange field, it is known that
in one dimension only chiral symmetry
can protect the existence of two degenerate bound states at the same end~\cite{Schnyder2008,kitaev2009periodic}.
We find that the chiral symmetry operator for the one-dimensional Hamiltonian
$\mathcal{H}(k_{x},0)$ has the form $\mathcal{C}=s_{z}\sigma_{y}$. Accordingly, one can define
a winding number to characterize the full Hamiltonian $\mathcal{H}(k_{x},0)$.
The winding number is given by~\cite{Ryu2010}
\begin{eqnarray}
\omega=\frac{1}{2\pi i}\int_{BZ}dk_{x}\text{Tr}[Q^{-1}(k_{x})\partial_{k_{x}}Q(k_{x})],
\end{eqnarray}
where ``BZ'' stands for Brillouin zone, ``Tr'' stands for the trace operation, and $Q(k_{x})$ is related to the Hamiltonian
and determined by rewriting the Hamiltonian into a new basis under which the chiral
symmetry operator is diagonal, i.e., $\mathcal{\tilde{C}}=diag\{1,1,-1,-1\}$, correspondingly,
\begin{eqnarray}
\tilde{\mathcal{H}}(k_{x},0)=\left(
                     \begin{array}{cc}
                       0 & Q(k_{x}) \\
                       Q^{\dag}(k_{x}) & 0 \\
                     \end{array}
                   \right).
\end{eqnarray}
Here the explicit form of $Q(k_{x})$ is
\begin{eqnarray}
Q(k_{x})=\left(
           \begin{array}{cc}
             M_{x}-iM_{y} & F(k_{x})-i\lambda_{\nu} \\
             F(k_{x})+i\lambda_{\nu} & M_{x}+iM_{y} \\
           \end{array}
         \right),
\end{eqnarray}
where
$F(k_{x})=t[2\cos(\frac{\sqrt{3}}{2}k_{x})+1]-i2\lambda_{\rm so}[\sin(\sqrt{3}k_{x})-2\sin(\frac{\sqrt{3}}{2}k_{x})]$.
Since the chiral symmetry is preserved even when $\lambda_{\nu}$, $M_{x}$ and $M_{y}$ are all nonzero,
the topological invariant will hold its value as long as the bulk energy gap of $\mathcal{H}(k_{x},0)$
remains open, therefore, the winding number can be determined by considering
the limit with $\lambda_{\nu}=M_{x}=M_{y}=0$. Accordingly, it is easy
to find that
\begin{eqnarray}
\omega=\frac{1}{2\pi i}\int_{BZ}dk_{x}2F^{-1}(k_{x})\partial_{k_{x}}F(k_{x})=2.
\end{eqnarray}
The chiral symmetry and the value of the winding number explain the robustness of the doubly-degenerate
crossing of the helical edge states on the armchair edges even when $\lambda_{\nu}\neq0$.

\begin{figure}[t]
\centering
\includegraphics[width=0.48\textwidth]{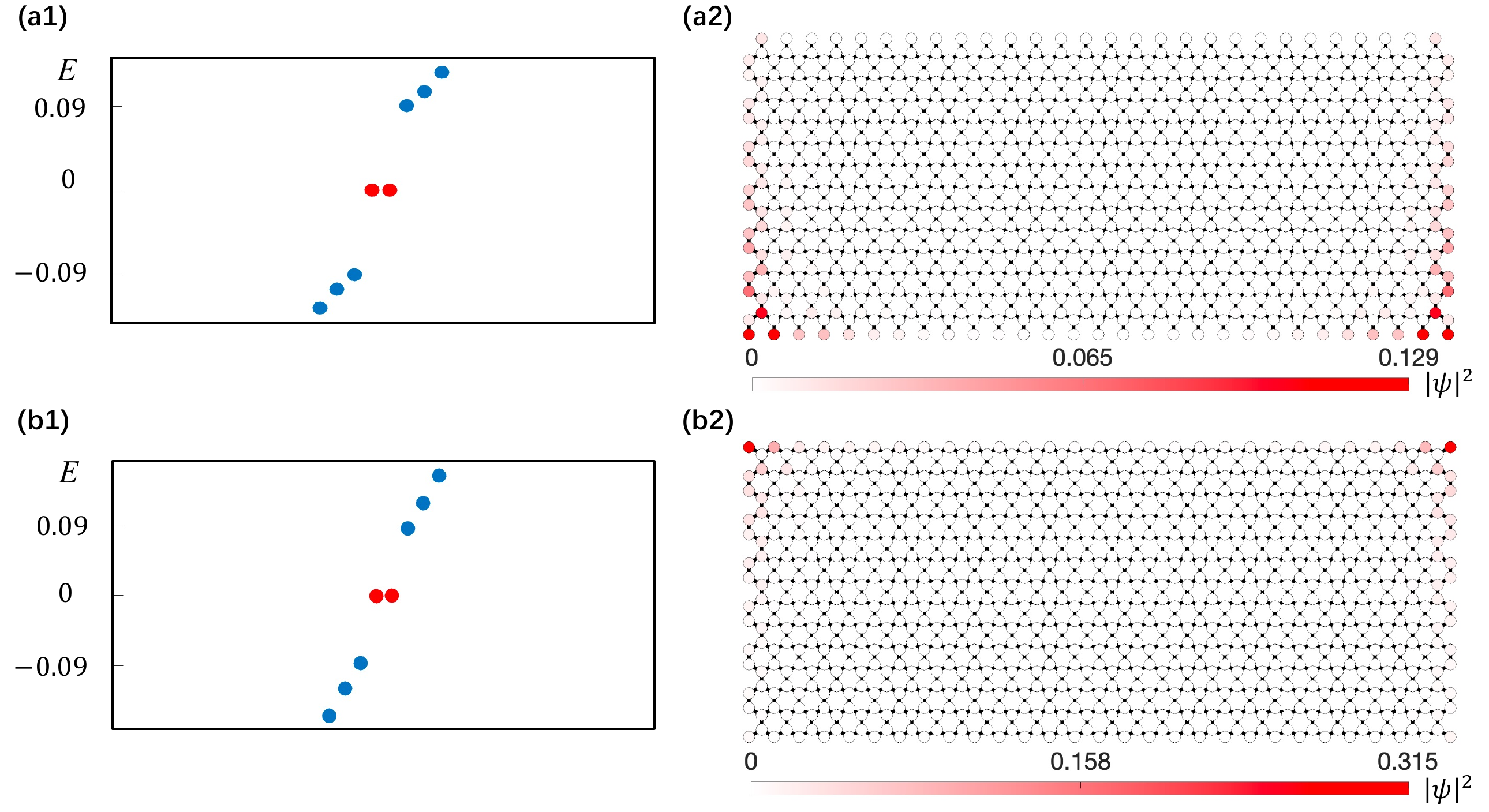}
\caption{(Color online) Corner bound states in a rectangular sample with
both $x$ and $y$ directions taking open boundary conditions. Chosen parameters are $t=1$, $\lambda_{\rm so}=0.1$,
$M_{x}=M_{y}=0.2$, $\gamma=0.3$, $\lambda_{\nu}=0$ and $N_{x}=N_{y}=28$.
The geometries of the samples are depicted in (a2) and
(b2), and a few corresponding eigenenergies near zero energy are shown in (a1) and (b1).
The red dots in (a1) and (b1) correspond to the eigenergies of the corner bound states.
The shade of the red color on the lattice sites in (a2) and (b2) reflects the weight of the probability
density of the corner bound states. }\label{corner}
\end{figure}

When $\gamma\neq1$, it is easy to find that the antiferromagnetic term in Hamiltonian (\ref{normal})
commutes with the chiral symmetry operator, i.e., $[(\bM\cdot \mathbf{s})\sigma_{z},\mathcal{C}]=0$,
indicating that the antiferromagnetic term breaks the chiral symmetry
of $\mathcal{H}(k_{x},0)$. As a result,  the protection of the crossing
at $k_{y}=0$ from the chiral
symmetry is lifted and the helical edge states
on the armchair edges would be gapped out by the antiferromagnetic term.
As shown in Figs.\ref{armchair}(c)(d), the numerical results confirm this expectation,
reflecting the correctness of our analysis.

The opening of an energy gap to the helical edge states on the armchair
edges implies an important consequence: one no longer needs to avoid
the armchair edges to achieve corner bound states. Now bound states are also possible
to emerge at the corners corresponding to the
intersections of armchair edges and other types of edges, such as zigzag or
beard edges. By numerical calculations, we confirm this expectation, as
shown in Fig.\ref{corner}. According to the numerical results shown in
Figs.\ref{corner}(a1)(a2), one can see that the two bound states are localized around
the two bottom corners for a rectangular sample with armchair edges in the $x$ direction and
beard edges in the $y$ direction. Interestingly, when the beard edges
are modified to zigzag edges by changing only the outermost sublattices of the $y$-normal edges,
the positions of the two bound states are found to shift dramatically
to the two top corners, as illustrated in  Figs.\ref{corner}(b1)(b2).
We will adopt the edge theory to show in the next section that this is because
the boundary Dirac mass has a sensitive dependence on the boundary
sublattice termination, and can switch its sign when the boundary
sublattice termination is changed from one type to the other.

It is worth pointing out that here the antiferromagnetic or ferrimagnetic exchange field
does not favor the realization of corner bound states.
This can be proved by numerical calculations or simply inferred by noting that in the limit $\lambda_{\nu}=0$ and $\gamma=-1$,
the momentum-independent antiferromagnetic exchange field anticommutes with all other terms
in the Hamiltonian (\ref{normal}). Since this implies that the antiferromagnetic term will introduce
a constant gap to the energy spectrum which cannot be closed by tuning the parameters of all other terms,
the resulting gapped phase is topologically connected to an atomic trivial insulator without any types of
topological mid-gap states on the boundary.

\section{Bound states at sublattice domain walls}
\label{sec4}

In the following, we are going to show that bound states can also be achieved even without
the existence of sharp corners, and their locations can be freely tuned
by taking advantage of the sublattice degrees of freedom.
In a previous work,  we have revealed that for the Kane-Mele model,
the boundary sublattice terminations have a strong impact on
the helical edge states, such as the shift of the boundary
Dirac points  (the crossing point of the energy spectra for helical edge states)
from one time-reversal invariant momentum to the other in the boundary
Brillouin zone~\cite{Zhu2021sublattice}. In addition, the sublattice terminations can also
strongly affect the boundary Dirac mass induced by superconductivity and hence
the formation of Dirac-mass domain walls supporting Majorana Kramers pairs~\cite{Zhu2021sublattice}.

\begin{figure}[t]
\centering
\includegraphics[width=0.48\textwidth]{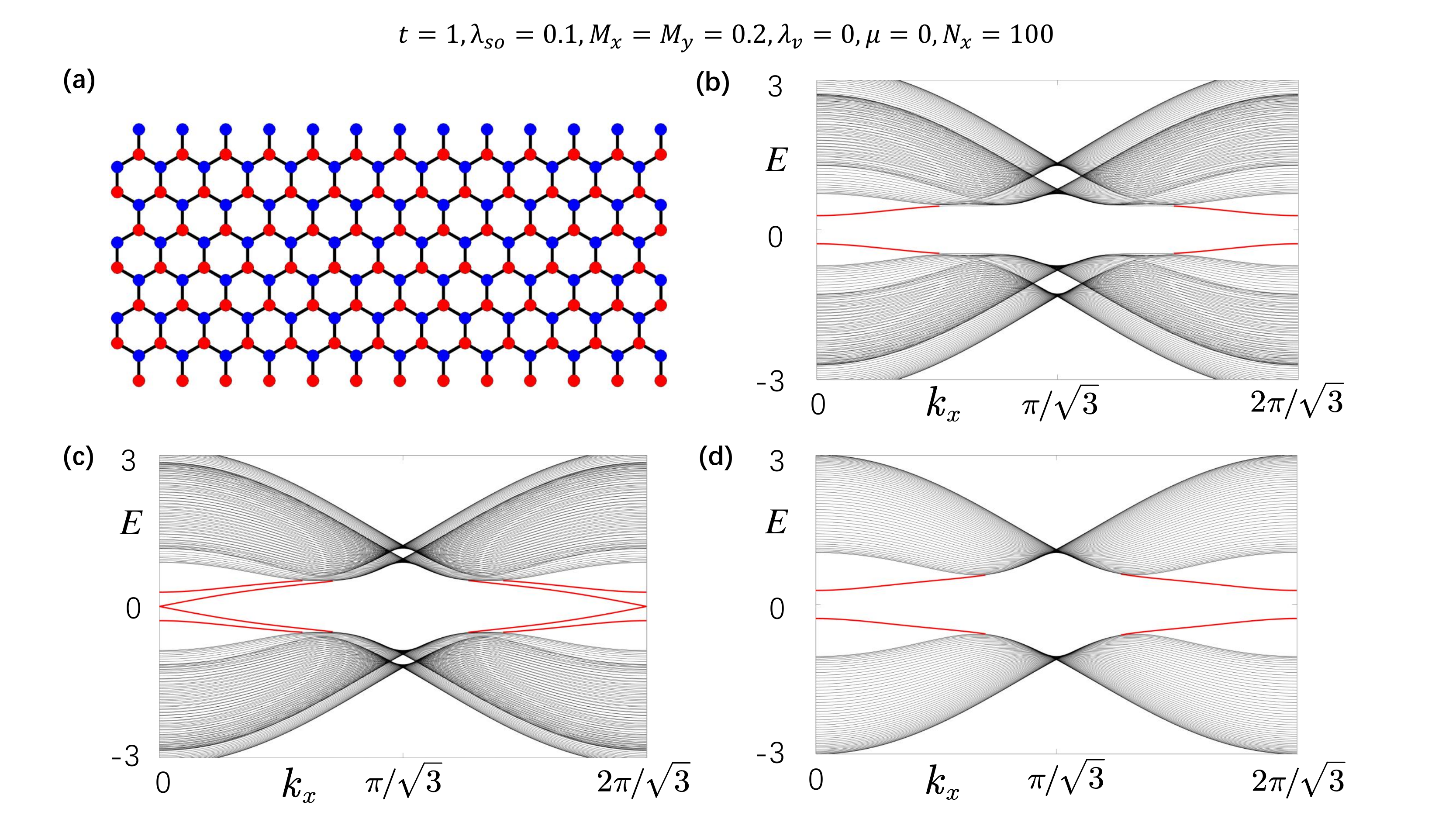}
\caption{(Color online) Energy spectra for a ribbon with
periodic boundary conditions in the $x$ direction and beard edges
in the $y$ direction.
Chosen parameters are $t=1$, $\lambda_{\rm so}=0.1$,
$M_{x}=M_{y}=0.2$, $\lambda_{\nu}=0$ and $N_{y}=100$.
(a) A schematic diagram  of a sample with beard edges in the $y$ direction.
Blue and red sites correspond to A-type and B-type sublattices, respectively.
The red solid lines in (b-d) denote energy spectra of the edge states on the beard edges.
(b) $\gamma=1$,
the helical edge states on the upper and lower beard edges are gapped out.
(c) $\gamma=0$, the helical edge states on the upper beard edge are gapped out, while
the helical edge states on the lower beard edge remain almost gapless. (d) $\gamma=-1$,
the helical edge states on the upper and lower beard edges are gapped out.}\label{beard}
\end{figure}

\begin{figure}[t]
\centering
\includegraphics[width=0.48\textwidth]{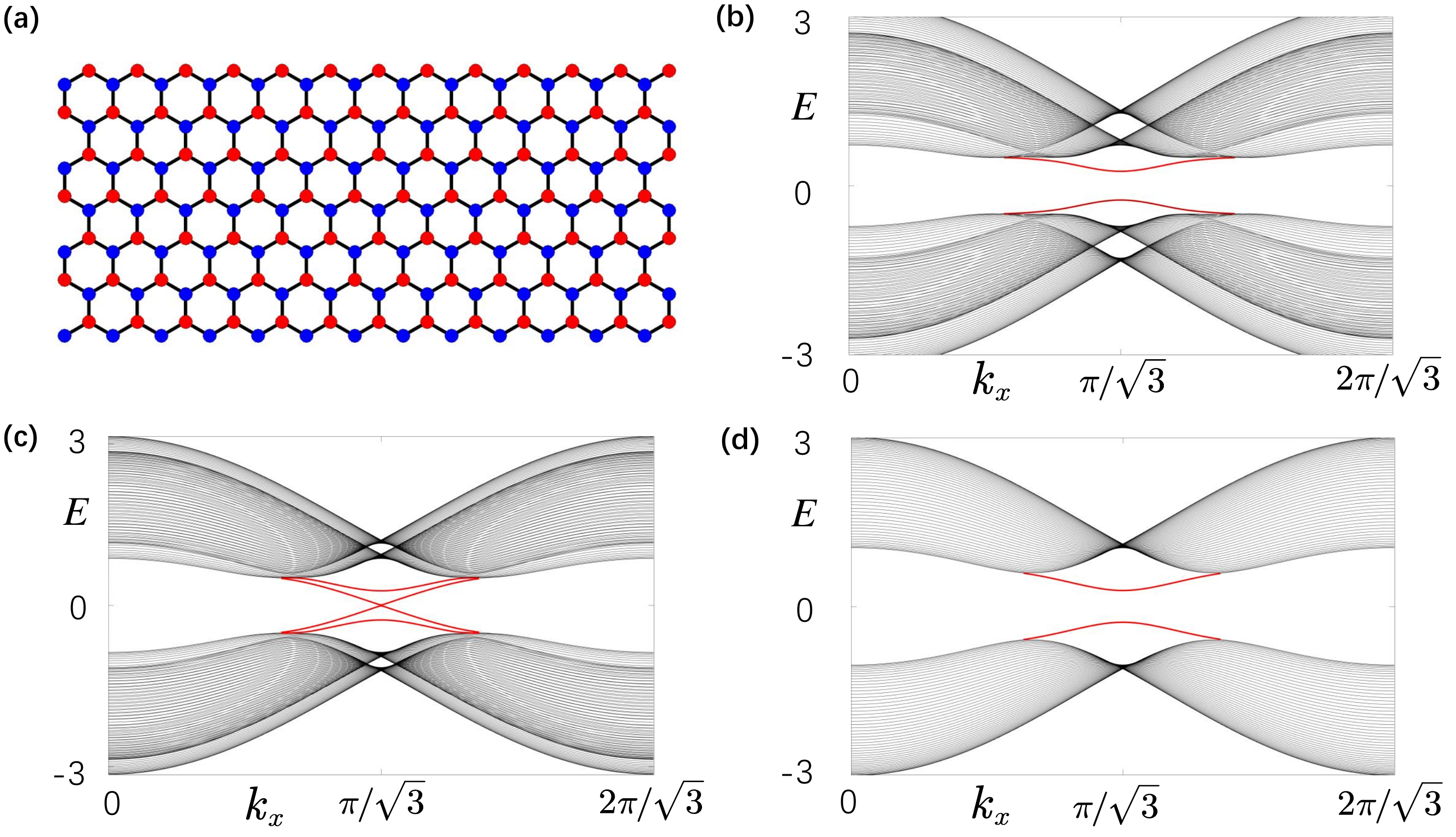}
\caption{(Color online) Energy spectra for a ribbon with
periodic boundary conditions in the $x$ direction and zigzag edges
in the $y$ direction.
Chosen parameters are $t=1$, $\lambda_{\rm so}=0.1$,
$M_{x}=M_{y}=0.2$, $\lambda_{\nu}=0$ and $N_{y}=100$. (a)
A schematic diagram  of a sample with zigzag edges in the $y$ direction.
The red solid
lines in (b-d) denote energy spectra of the edge states on the zigzag edges.
(b) $\gamma=1$, the helical edge states on the upper and lower zigzag edges are gapped out.
(c) $\gamma=0$, the helical edge states on the upper zigzag edge remain almost gapless, while
the helical edge states on the lower zigzag edge are gapped out. (d) $\gamma=-1$,
the helical edge states on the upper and lower zigzag edges are gapped out.  }\label{zigzag}
\end{figure}

To explore the impact of sublattice terminations on the boundary Dirac mass induced by
exchange field, we first numerically calculate the energy spectra for a ribbon
with the $x$ direction taking periodic boundary conditions and the $y$ direction
having only beard or zigzag edges, as illustrated in Figs.\ref{beard}(a) and \ref{zigzag}(a).
According to the results presented in Figs.\ref{beard}(b-d) and \ref{zigzag}(b-d),
one can infer that the boundary energy spectra (red solid lines) for
the upper and lower beard or zigzag edges are degenerate when
$\gamma=\pm1$, suggesting that the Dirac masses induced by the exchange field
on the upper and lower beard or zigzag edges have the same magnitude for these two
limiting cases. On the other hand, the results for $\gamma=0$ clearly reveal that the boundary Dirac mass strongly depends on
the sublattice termination for a given boundary. To be specific, let us focus on
the upper $y$-normal boundary for a more detailed discussion. When $\gamma=0$, based on the edges at which
the mid-gap states are localized, we find that the Dirac mass
is vanishingly small when the upper $y$-normal boundary terminates with B sublattices, as illustrated
in Fig.\ref{zigzag}(c).
Obviously, the smallness of the Dirac mass should be related to the fact that exchange fields on B sublattices are absent
when $\gamma=0$. This implies that, for a given sublattice termination,
the magnitude of the exchange field on the corresponding sublattices determines
the main contribution to the magnitude of the Dirac mass.

As it turns out that both ferromagnetic and antiferromagnetic exchange fields can induce a finite Dirac mass
to the helical edge states, regardless of the boundary sublattice terminations, a natural question to ask is:
for a given type of exchange field, do the boundary Dirac masses associated with the two kinds of
sublattice terminations
have the same sign or opposite signs?
Naively, one may think that when $\gamma>0$, since the exchange fields take
the same direction on the two types of sublattices, the Dirac masses  should also take the same sign
for the two kinds of sublattice terminations. In contrast, when $\gamma<0$,
since the exchange fields take
opposite directions on the two kinds of sublattices, one may think that
the Dirac masses associated with the two kinds of sublattice terminations
should also take opposite signs. However, we find that the results are just
the opposite. To show this, we focus on the upper $y$-normal boundary and derive
the low-energy Hamiltonians describing the boundary physics
on the zigzag and beard edges.

Here we focus on the limit $\lambda_{\nu}=0$ and
consider $t$ and $\lambda_{\rm so}$ to be positive constants for
the convenience of discussion. First, let us consider
the upper $y$-normal boundary to be a beard edge (terminating with A sublattices, see the upper
edge in Fig.\ref{beard}(a)).
We find that the corresponding low-energy Hamiltonian has the form
(see details in Appendix A)
\begin{eqnarray}
\mathcal{H}_{b}(q_{x})=vq_{x}s_{z}+M_{x}s_{x}+M_{y}s_{y},\label{lowbeard}
\end{eqnarray}
where $q_{x}$ denotes a small momentum measured from $k_{x}=0$ at which the
boundary Dirac point is located (see the dispersion of edge states in Fig.\ref{beard}), and the velocity of
the helical edge states is $v=3\sqrt{3}\lambda_{\rm so}$.
On the other hand, when the upper $y$-normal boundary is a zigzag edge (terminating with B sublattices, see the upper
edge Fig.\ref{zigzag}(a)),
we find that
the low-energy Hamiltonian has the form (see details in Appendix B)
\begin{eqnarray}
\mathcal{H}_{z}(q_{x}')=v'q_{x}'s_{z}+\frac{1-\gamma\eta^{2}}{1+\eta^{2}}(M_{x}s_{x}+M_{y}s_{y}),\label{lowzigzag}
\end{eqnarray}
where $q_{x}'$ denotes the momentum measured from $k_{x}=\pi/\sqrt{3}$ at which the
boundary Dirac point is located (see Fig.\ref{zigzag}(c)), and the explicit
expressions of the parameters read
\begin{eqnarray}
v'&=&\frac{2\sqrt{3}t\lambda_{\rm so}\eta}{1+\eta^{2}}+\frac{2\sqrt{3}\lambda_{\rm so}(\eta^{2}-1)}{1+\eta^{2}},\nonumber\\
\eta&=&\frac{4t\lambda_{\rm so}}{\sqrt{t^{2}(t^{2}+16\lambda_{\rm so}^{2})}-t^{2}}.
\end{eqnarray}
In real materials, $\lambda_{\rm so}$ is commonly much smaller than $t$. Focusing
on the regime $\lambda_{\rm so}\ll t$, it is easy to find that $\eta\simeq t/2\lambda_{\rm so}\gg1$, and
the Hamiltonian can be approximately reduced as
\begin{eqnarray}
\mathcal{H}_{z}(q_{x}')\approx 2vq_{x}'s_{z}-\gamma(M_{x}s_{x}+M_{y}s_{y}).
\end{eqnarray}
The two low-energy Hamiltonians, (\ref{lowbeard}) and (\ref{lowzigzag}),
 provide a clear understanding of the
dependence of the boundary Dirac mass on the exchange field and
sublattice termination. Remarkably, the results show that the Dirac masses
on the zigzag and beard edges with the same orientation have opposite signs when $\gamma>\gamma_{c}=\eta^{-2}$,
and the same sign when $\gamma<\gamma_{c}$. When $\eta\gg1$,
the critical value $\gamma_{c}$ can be set as zero.

\begin{figure}[t]
\centering
\includegraphics[width=0.48\textwidth]{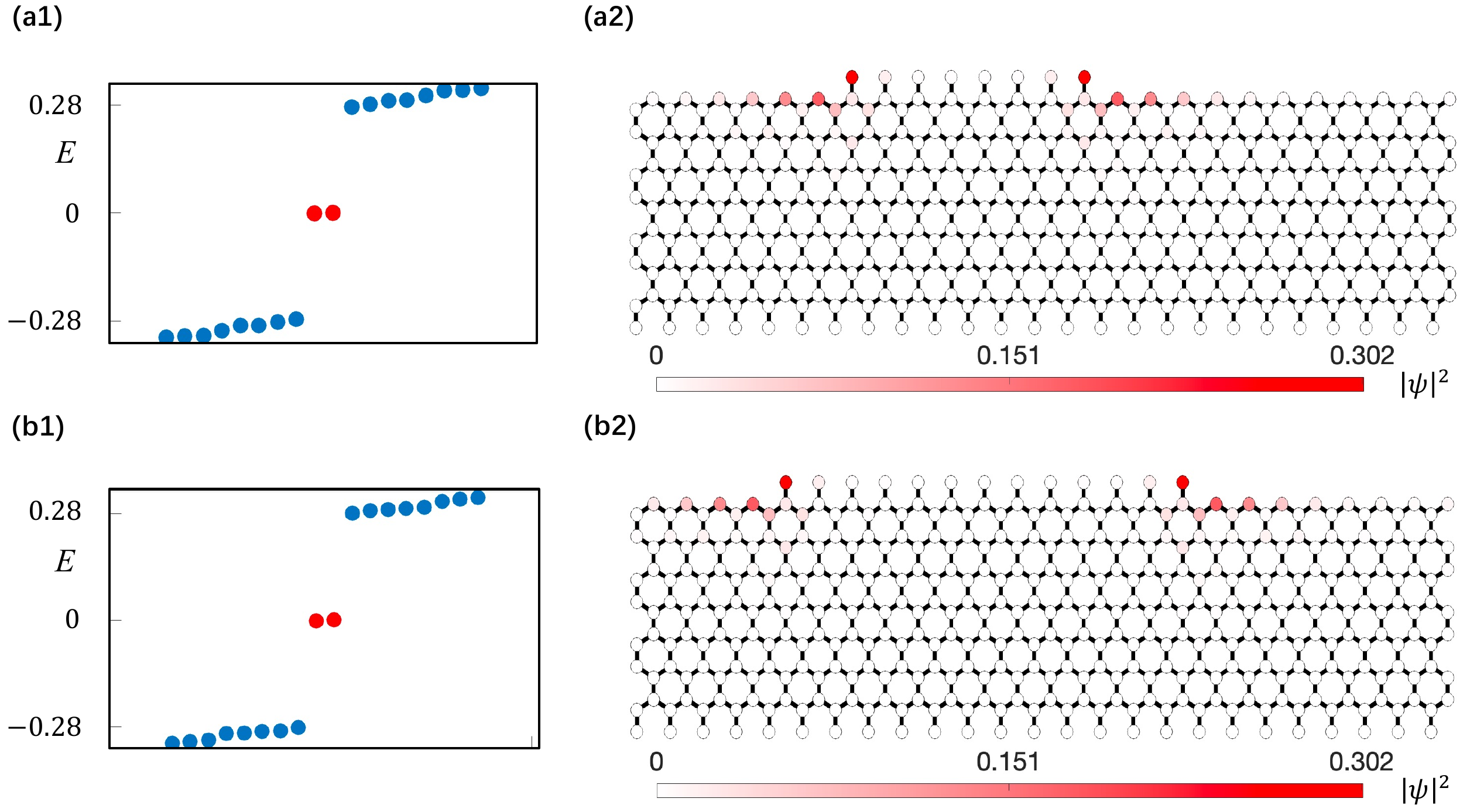}
\caption{(Color online) Tunable bound states at the sublattice domain walls.
Chosen parameters are $t=1$, $\lambda_{\rm so}=0.1$,
$M_{x}=M_{y}=0.2$, $\gamma=1$ and $\lambda_{\nu}=0$. (a1) and (b1) show the energy spectra (only a few eigenenergies
near zero energy are shown) corresponding to systems with the geometries shown in
(a2) and (b2), respectively. Periodic boundary conditions are imposed in the $x$
direction, namely, the left and right armchair edges in (a2) and (b2) are connected
when diagonalizing the Hamiltonian. The red dots in (a1) and (b1) correspond to the eigenenergies of the
bound states at the sublattice domain walls.
The shade of the red color on the lattice sites in (a2) and (b2) reflects the weight of the probability
density of the bound states. }\label{sdw}
\end{figure}

As Dirac masses of opposite signs lead to the formation of
domain walls hosting bound states~\cite{jackiw1976b}, apparently, the dependence of
Dirac mass on the sublattice termination shown in the two
low-energy boundary Hamiltonians suggests that Dirac-mass domain
walls can form on the same $y$-normal boundary. Put it more explicitly,
when $\gamma>\gamma_{c}$ and the $y$-normal boundary consists of two flat parts,
with one part taking the beard edge (terminating with A sublattices) and
the other  taking the zigzag edge (terminating with B sublattices),
then the sublattice domain walls, which correspond to the intersections of
the beard and zigzag edges, are Dirac-mass domain walls hosting bound states.
The numerical results shown in Fig.\ref{sdw} confirm this expectation.  We would like
to emphasize two important properties of the sublattice domain walls
that can be inferred from the numerical results.
First, as the sublattice domain walls on the same boundary can support
bound states, it suggests that sharp corners are not a necessary condition
to achieve bound states in a second-order topological phase if the boundary
Dirac mass shows sensitive dependence on the sublattice terminations. Indeed,
Fig.\ref{sdw} demonstrates that bound states are present even though there
is no sharp corner in the system with periodic boundary conditions in one direction. Second,
as the bound states are associated with the sublattice domain walls, it suggests that
the locations of the bound states can be tuned by locally manipulating
the sublattice termination. This fact can be intuitively inferred by a comparison of the
locations of the bound states in Figs.\ref{sdw}(a2)(b2).

In addition to the exchange field, it is known that the superconductivity can also
induce a Dirac mass to the helical edge states and gap them out~\cite{Fu2009qshe}. An important fact to note is that
 the Dirac masses induced
by exchange field and superconductivity are competing in nature. Above we have shown
that the Dirac masses induced by exchange field on the two sides of a sublattice domain wall
can have different magnitude and signs. Apparently, this raises the possibility
to realize domain walls with the Dirac mass on one side dominated by the superconductivity and
on the other side dominated by the exchange field. It is known that MZMs will emerge
at such domain walls~\cite{Fu2009qshe,Yan2019hoscb,Jack2019observation,Wu2020SOTSC}. In the following, we consider conventional s-wave superconductivity to demonstrate
that MZMs can be realized at the sublattice domain walls.

\section{Majorana zero modes at sublattice domain walls}
\label{sec5}

Before proceeding, it is worth pointing out that a number of proposals on the realization of
2D second-order TSCs or topological superfluids have
been raised in the past few years, including TI/superconductor heterostructures~\cite{Yan2018hosc,Wang2018hosc,
Liu2018hosc,Zhang2019hoscb,Pan2019SOTSC,Wu2020SOTSC,Majid2020hosca,Laubscher2020mcm,Li2021bts,Li2021mcm,Tan2022corner,Wu2022corner},
superconductors with mixed-parity pairings~\cite{Wang2018weak,Wu2019hosc,Ikegaya2021mcm,Roy2020HOTSC},
spin-orbit coupled superconductors with s+id pairing~\cite{Zhu2019mixed,Majid2020hoscb},
odd-parity superconductors~\cite{Zhu2018hosc,Yan2019hosca,Ahn2020hosc,Hsu2020hosc,Li2022hosc,Scammell2022},
etc~\cite{Zeng2019mcm,Volpez2019SOTSC,Franca2019SOTSC,Huang2019mirror,wu2020boundaryobstructedb,Wu2020cornercold,Wu2021cornercold,
Qin2022hosc,Roy2021HOTSC},
also with the MZMs localized at sharp corners
being the smoking gun.  Among the various proposals, the TI/superconductor heterostructures
are arguably most close to implementation owing to the abundance of candidate materials.

By putting the TI described by the Kane-Mele model in proximity
to an s-wave superconductor, the whole system can
be effectively described by a Bogoliubov-de Gennes (BdG) Hamiltonian. Consider
the basis $\Psi_{\bk}=(\psi_{\bk},\psi_{-\bk}^{\dag})^{T}$,
$H=\frac{1}{2}\sum_{\bk}\Psi_{\bk}^{\dag}\mathcal{H}_{\rm BdG}(\bk)\Psi_{\bk}$ with
\begin{eqnarray}
\mathcal{H}_{\rm BdG}(\bk)=\left(
                   \begin{array}{cc}
                     \mathcal{H}(\bk)-\mu s_{0}\sigma_{0} & i\Delta s_{y}\sigma_{0} \\
                     -i\Delta^{*} s_{y}\sigma_{0} & -\mathcal{H}^{*}(-\bk)+\mu s_{0}\sigma_{0} \\
                   \end{array}
                 \right),\quad
\end{eqnarray}
where $\mu$ is the chemical potential, and $\Delta$ is the s-wave pairing amplitude.
Below we will assume $\Delta$ to be a momentum-independent real constant for the convenience
of discussion.

To  show intuitively that MZMs can emerge at the sublattice domain walls, we derive  the corresponding low-energy boundary Hamiltonian
 based on the BdG Hamiltonian. For generality, now we consider
the staggered sublattice potential to be finite.  Also focusing on the upper $y$-normal boundary for illustration
of the key physics, we find that, for the beard edge,
the low-energy boundary Hamiltonian reads (see details in Appendix A)
\begin{eqnarray}
\mathcal{H}_{\rm BdG;b}(q_{x})& = & vq_{x} \tau_{0}s_z
   +M_{x} \tau_{z} s_{x}+M_y \tau_{0}s_{y}\nonumber\\
   &&-(\mu - \lambda_{\nu}) \tau_{z}s_{0}-\Delta\tau_{y}s_{y},\label{BdGbeard}
\end{eqnarray}
and for the zigzag edge, the low-energy boundary Hamiltonian reads (see details in Appendix B)
\begin{eqnarray}
\mathcal{H}_{\rm BdG;z}(q_{x}')& = & v'q_{x}' \tau_{0}s_z
   +\frac{1-\gamma\eta^{2}}{1+\eta^{2}}(M_{x} \tau_{z} s_{x}+M_{y}\tau_{0}s_{y})\nonumber\\
   &&-(\mu +\frac{\eta^{2}-1}{1+\eta^{2}}\lambda_{\nu}) \tau_{z}s_{0}-\Delta\tau_{y}s_{y},\label{BdGzigzag}
\end{eqnarray}
where $\tau_{0}$ and $\tau_{x,y,z}$ are identity matrix and Pauli matrices in the particle-hole
space. It is readily seen that the staggered potential effectively induces an opposite shift in
the chemical potential. This is easy to understand since the terminating sublattices for
these two kinds of edges are different and hence have
different potentials. As will be shown below, this can benefit the realization of MZMs
at the sublattice domain walls.

\begin{figure}[t]
\centering
\includegraphics[width=0.48\textwidth]{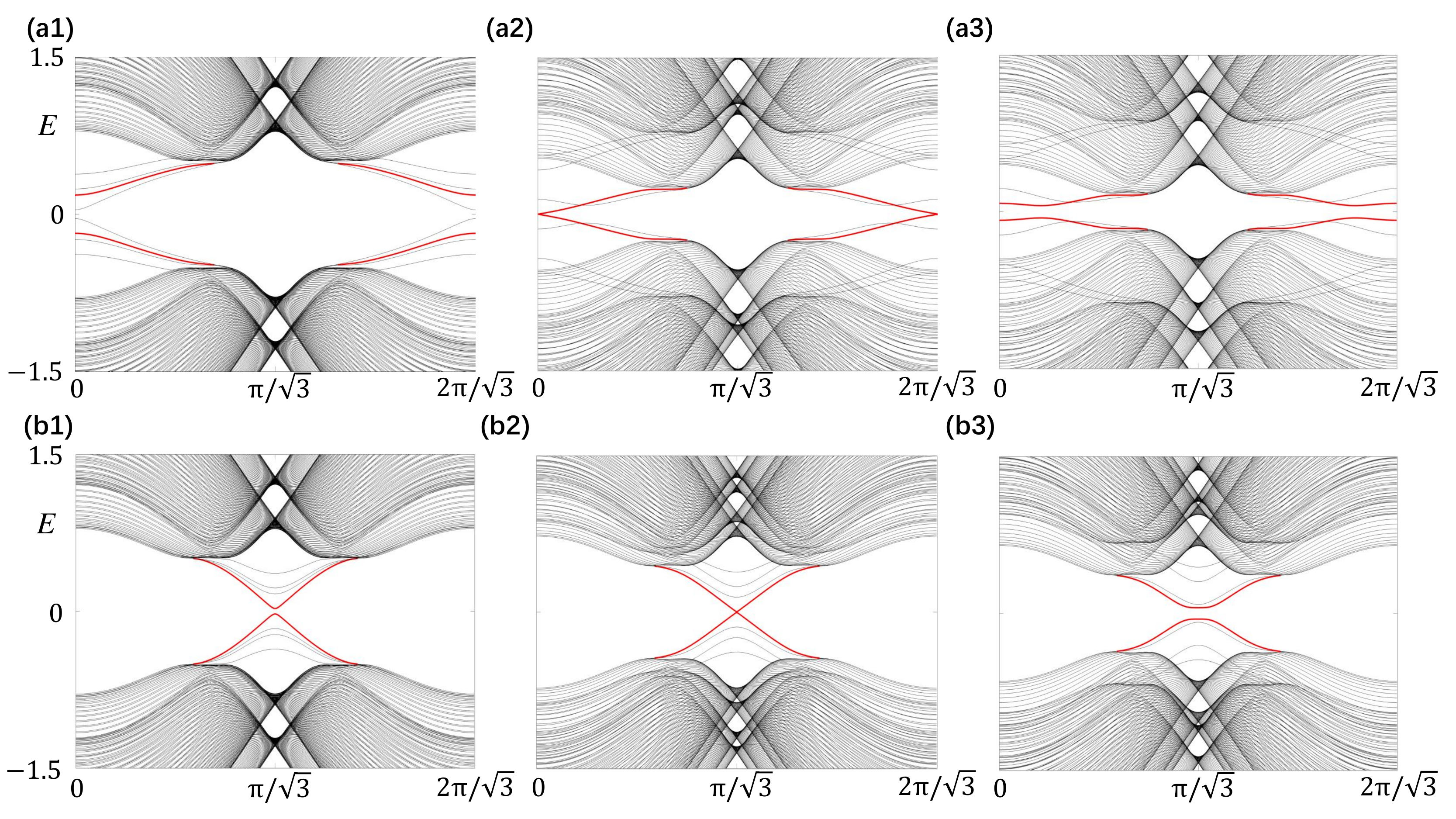}
\caption{(Color online) The evolution of boundary energy gap on the upper
$y$-normal edge with
respect to $\mu$ for a cylindrical geometry  with periodic
boundary conditions in the $x$ direction.
Chosen parameters are $t=1$, $\lambda_{\rm so}=0.1$,
$M_{x}=M_{y}=0.2$, $\gamma=0.5$, $\lambda_{\nu}=0$ and $\Delta=0.1$. For (a1)-(a3), the open boundaries are beard edges,
and the critical value of $\mu$ at which the boundary energy gap of the upper edge
is equal to $0.265$ according to the chosen parameters.
For (b1)-(b3), the open boundaries are zigzag edges, and the critical value of
$\mu$ is equal to $0.077$. (a1) $\mu=0$,
(a2) $\mu=0.265$, (a3) $\mu=0.35$, (b1) $\mu=0$, (b2) $\mu=0.07$, and (b3) $\mu=0.15$. }\label{boundary}
\end{figure}

Without loss of generality, let us still focus on the regime $t\gg\lambda_{\rm so}$ so that $\eta\gg1$ and the form of the low-energy
boundary Hamiltonian on the zigzag edge can be simplified as
\begin{eqnarray}
\mathcal{H}_{\rm BdG;z}(q_{x}')& \approx & v'q_{x}' \tau_{0}s_z
   -\gamma(M_{x} \tau_{z} s_{x}+M_y\tau_{0}s_{y})\nonumber\\
   &&-(\mu +\lambda_{\nu}) \tau_{z}s_{0}-\Delta\tau_{y}s_{y}.\label{AH}
\end{eqnarray}
It is straightforward to find that the gap-closing condition of the boundary
energy spectrum for the beard edge is
\begin{eqnarray}
M=\sqrt{(\mu-\lambda_{\nu})^{2}+\Delta^{2}},
\end{eqnarray}
and for the zigzag edge, the gap-closing condition is
\begin{eqnarray}
|\gamma|M=\sqrt{(\mu+\lambda_{\nu})^{2}+\Delta^{2}},\label{zcondition}
\end{eqnarray}
where $M=\sqrt{M_{x}^{2}+M_{y}^{2}}$. It is worth noting
that, for simplicity, the gap-closing condition for the zigzag edge
is obtained via the approximate Hamiltonian (\ref{AH}). The accurate
condition can also be easily obtained according to the Hamiltonian (\ref{BdGzigzag}),
but will have a somewhat more complex expression (see Eq.(\ref{accurate})).
In Fig.\ref{boundary},
we assume that the exchange field is fixed,  and show the evolution of the boundary energy spectra (red solid lines) with respect
to $\mu$. We find that the critical $\mu_{c}$ at which the boundary energy gap
on the upper edge gets closed agrees excellently with the value
predicted by the low-energy boundary Hamiltonians (\ref{BdGbeard}) and (\ref{BdGzigzag}),
reflecting the power of the edge theory in describing the  boundary physics.

For a given edge, the gap closure of the
boundary energy spectrum signals a change of the boundary topology.
For the upper beard edge, its Dirac mass falls into the superconductivity-dominated region when
$M<M_{c,b}\equiv\sqrt{(\mu-\lambda_{\nu})^{2}+\Delta^{2}}$, and the exchange-field-dominated region
when $M>M_{c,b}$. Similarly, the Dirac mass of the upper zigzag edge falls into
the superconductivity-dominated region when
$|\gamma| M<M_{c,z}\equiv\sqrt{(\mu+\lambda_{\nu})^{2}+\Delta^{2}}$, and
the exchange-field-dominated region when $|\gamma|M>M_{c,z}$. When the Dirac masses on the upper beard and zigzag
edges fall into different regions,
the sublattice domain walls will bind MZMs.

Without loss of generality, let us consider $\mu\geq0$ and $\lambda_{\nu}\geq0$ to exemplify
the physics. With this choice,  the condition to realize MZMs at the
sublattice domain walls is
\begin{eqnarray}
M>M_{c,b}, \quad |\gamma|M<M_{c,z}.\label{Maincriterion}
\end{eqnarray}
Since the two inequalities above are independent of the sign of $\gamma$,
it indicates that MZMs at sublattice domain walls can be achieved
for both ferromagnetic and antiferromagnetic exchange fields as long as the two inequalities
are simultaneously fulfilled.

\begin{figure}[t]
\centering
\includegraphics[width=0.48\textwidth]{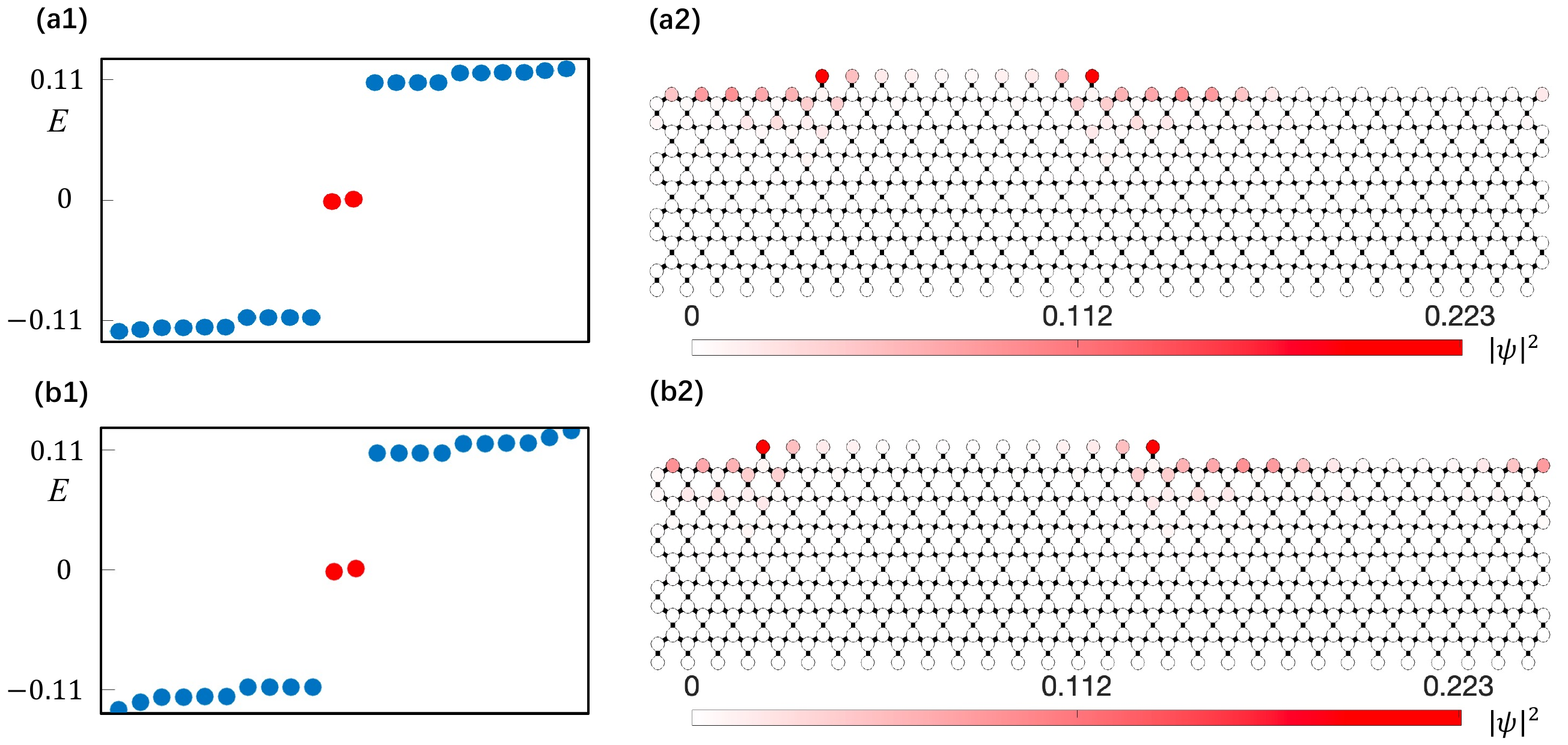}
\caption{(Color online) Tunable MZMs at sublattice domain walls
for a ferromagnetic exchange field. Chosen parameters are $t=1$, $\lambda_{\rm so}=0.1$,
$M_{x}=M_{y}=0.2$, $\gamma=1$, $\Delta=0.15$, $\lambda_{\nu}=0.2$, and $\mu=0.2$.
The geometry considered is depicted in (a2) and
(b2), and a few corresponding eigenenergies near zero energy are shown in (a1) and (b1). Periodic boundary conditions are
imposed in the $x$ direction.
The red dots in the energy spectra denote MZMs (their energies are not exactly
zero due to splitting induced by finite-size effects)
at the sublattice domain walls. The right panels show their probability density
profiles with  the shade of the red color
on the lattice sites reflecting the weight. }\label{Fexchange}
\end{figure}

\begin{figure}[t]
\centering
\includegraphics[width=0.48\textwidth]{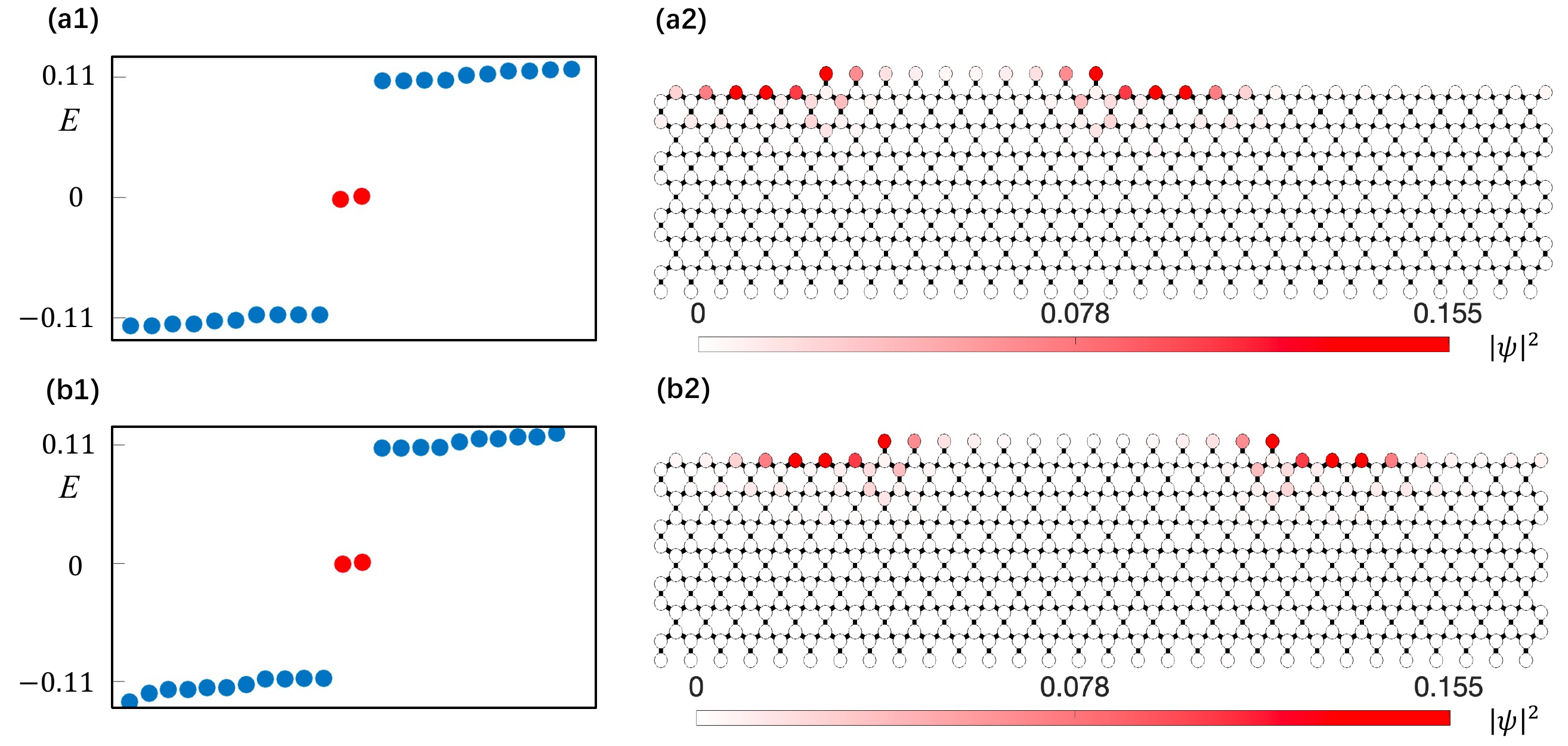}
\caption{(Color online) Tunable MZMs at sublattice domain walls
for an antiferromagnetic exchange field. All parameters are the same as in Fig.\ref{Fexchange}
except $\gamma=-1$. }\label{AFexchange}
\end{figure}

We would like to make a further remark on Eq.(\ref{Maincriterion}). The two inequalities suggest that the
topological region supporting MZMs can be made very sizable.
For instance, by tuning $\mu=\lambda_{\nu}$, $\gamma=0$, MZMs
can be realized once $M>|\Delta|$. In Fig.\ref{Fexchange}, we consider
a ferromagnetic exchange field and show
the realization of MZMs at the sublattice domain walls for a sample
with cylindrical geometry (left and right edges are connected, hence there is no sharp
corner). By
a comparison of Fig.\ref{Fexchange}(a2) and Fig.\ref{Fexchange}(b2),
it is easy to see that the positions of MZMs can be tuned to any place
on the upper edge by manipulating the boundary sublattice terminations.
Apparently, one can also manipulate two sublattice domain walls to move
toward each other, then one can study the splitting and annihilation of two
MZMs. In Fig.\ref{AFexchange}, we consider an antiferromagnetic exchange field and
show explicitly that the physics is similar.

\section{Discussion and Conclusion}
\label{sec6}

In this paper, we have shown when the lattice structure has sublattice degrees of freedom,
the bound states in second-order TIs and
TSCs are unnecessarily pinned at some specific sharp corners. By adjusting the boundary
sublattice terminations to form sublattice domain walls, we have shown that the positions of the bound states
can be freely manipulated. For the honeycomb lattice considered, if one
designs a sample with the diamond shape as considered in Ref.\cite{Ren2020corner} or
also with the honeycomb shape so that all edges take either
the beard-type or the zigzag-type sublattice termination, then
the sublattice domain walls allow to form at any place on the boundary.
Accordingly, the positions of the bound states
can be manipulated to any place on the boundary.
It is reasonable to expect that such a sublattice-enriched tunability
would benefit the manipulation and application of the bound states, e.g.,
braiding  MZMs~\cite{Zhang2020SOTSCa,Zhang2020SOTSCb,Pahomi2020}.

About the experimental implementation, we would like
to first emphasize that our predictions are
relevant to both quantum materials and classical systems.
For quantum-material realization, one route is to
apply a magnetic field to a two-dimensional first-order TI described by the Kane-Mele model, such as silicene, germanene,
stanene~\cite{Liu2011silicenea,Liu2011siliceneb,Ezawa2012silicene}.
As twisted transition metal dichalcogenide homobilayers are predicted to effectively realize
Kane-Mele model and allow various types of magnetic orders~\cite{Wu2019twisted,Zare2021twisted}, they may also serve
as a platform to explore the predicted boundary physics.
Another route is to find intrinsic magnetic second-order TI with sublattice degrees of freedom through first-principle calculations~\cite{Chen2020corner,Luo2022corner,Mu2022Fese}.
Generalizations to the superconducting counterpart can be simply achieved by
putting the above two classes of systems in proximity to a superconductor~\cite{Yan2019hoscb,Wu2020SOTSC}, as
demonstrated in Sect.\ref{sec5}. In quantum materials, as the lattice constant
is at the atomic length scale, adjusting boundary sublattice
terminations requires sophisticated tools, e.g., scanning tunneling microscope or scanning force
microscope~\cite{Stroscio1991,Custance2009}.
For classical-system realization, since the Kane-Mele model
subjected to an in-plane Zeeman field has been effectively realized
in an acoustic system\cite{Huang2022KM}, our prediction
on the realization of bound states at sublattice domain walls in a
second-order TI can be immediately explored.
It is worth emphasizing that the manipulation of bound states at sublattice
domain walls is expected to be much easier in classical systems than in
quantum materials due to  their much larger
length scales. For instance, one can simply remove one sublattice on the boundary
in an electric circuit by just removing all wires connected to that sublattice.
As a final remark, it is worth pointing out that since the wave functions of
the helical edge states decay exponentially away from the boundary, the boundary Dirac masses
are mainly contributed by the exchange field and superconductivity at the neighborhood of
the edges. In other words, the predicted boundary physics in this work can also
be realized even when the exchange field and superconductivity are nonuniform
or only appear at the neighborhood of the edges.

In summary, we have shown that the sublattice degrees of freedom and second-order
topology have an interesting interplay, which can lead to the presence of rich boundary physics,
such as the formation of highly controllable bound states.

\section{Acknowledgements}

D.Z. and Z.Y. are supported by the National Natural Science Foundation of China (Grant No.11904417
and No. 12174455) and the Natural Science Foundation of Guangdong Province
(Grant No. 2021B1515020026). M.Kh is supported
by the NSERC of Canada.

\appendix

\section{Low-energy boundary Hamiltonian on the beard edge}

The low-energy boundary Hamiltonians on the beard and zigzag edges for
the Kane-Mele model have been derived in a previous work of us~\cite{Zhu2021sublattice}, but
there we did not consider the staggered potential and exchange field.
Here for self-consistency, we provide the main steps of the derivation.

Start with the BdG Hamiltonian in the momentum space,
\begin{eqnarray}
\mathcal{H}_{\rm BdG}(\bk)&=&t(2\cos\frac{\sqrt{3}k_{x}}{2}\cos\frac{k_{y}}{2}+\cos k_{y})\tau_{z}s_{0}\sigma_{x}\nonumber\\
&&-t(2\cos\frac{\sqrt{3}k_{x}}{2}\sin\frac{k_{y}}{2}-\sin k_{y})\tau_{z}s_{0}\sigma_{y}\nonumber\\
&&+2\lambda_{\rm so}(\sin\sqrt{3}k_{x}-2\sin\frac{\sqrt{3}k_{x}}{2}\cos\frac{3k_{y}}{2})\tau_{0}s_{z}\sigma_{z}\nonumber\\
&&+\lambda_{\nu}\tau_{z}s_{0}\sigma_{z}-\mu\tau_{z}s_{0}\sigma_{0}\nonumber\\
&&+\frac{1+\gamma}{2}(M_{x}\tau_{z}s_{x}+M_{y}\tau_{0}s_{y})\sigma_{0}\nonumber\\
&&+\frac{1-\gamma}{2}(M_{x}\tau_{z}s_{x}+M_{y}\tau_{0}s_{y})\sigma_{z}\nonumber\\
&&-\Delta\tau_{y}s_{y}\sigma_{0},\label{BdGH}
\end{eqnarray}
where $\tau_{i}$, $s_{i}$ and $\sigma_{i}$ are Pauli matrices acting on
the particle-hole, spin and sublattice degrees of freedom, respectively, and
$\tau_{0}$, $s_{0}$ and $\sigma_{0}$ denote identity matrices in
the respective subspaces. For notational simplicity, the nearest-neighbor
lattice constant has been set to unity.

When the upper boundary is a beard edge, the numerical results show
that the corresponding boundary Dirac point is located at the time-reversal invariant momentum
 $k_{x}=0$ in the reduced boundary Brillouin zone. To derive
the low-energy boundary Hamiltonian in an analytical way,   we expand the bulk Hamiltonian around $k_{x}=0$ up to
the linear order in momentum (the expansion is only performed in the $k_{x}$ direction), leading to
\begin{eqnarray}
\mathcal{H}_{\rm BdG}(q_{x},k_{y})&=&t(2\cos\frac{k_{y}}{2}+\cos k_{y})\tau_{z}s_{0}\sigma_{x}\nonumber\\
&&-t(2\sin\frac{k_{y}}{2}-\sin k_{y})\tau_{z}s_{0}\sigma_{y}\nonumber\\
&&+2\sqrt{3}\lambda_{\rm so}q_{x}(1-\cos\frac{3k_{y}}{2})\tau_{0}s_{z}\sigma_{z}\nonumber\\
&&+\lambda_{\nu}\tau_{z}s_{0}\sigma_{z}-\mu\tau_{z}s_{0}\sigma_{0}\nonumber\\
&&+\frac{1+\gamma}{2}(M_{x}\tau_{z}s_{x}+M_{y}\tau_{0}s_{y})\sigma_{0}\nonumber\\
&&+\frac{1-\gamma}{2}(M_{x}\tau_{z}s_{x}+M_{y}\tau_{0}s_{y})\sigma_{z}\nonumber\\
&&-\Delta\tau_{y}s_{y}\sigma_{0},
\end{eqnarray}
where $q_{x}$ denotes a small momentum which is measured from $k_{x}=0$.
In the next step, we decompose the Hamiltonian into two parts, $\mathcal{H}_{\rm BdG}=\mathcal{H}_{0}+\mathcal{H}_{1}$,
with
\begin{eqnarray}
\mathcal{H}_{0}(q_{x},k_{y})&=&t(2\cos\frac{k_{y}}{2}+\cos k_{y})\tau_{z}s_{0}\sigma_{x}\nonumber\\
&&-t(2\sin\frac{k_{y}}{2}-\sin k_{y})\tau_{z}s_{0}\sigma_{y}, \nonumber\\
\mathcal{H}_{1}(q_{x},k_{y})&=&2\sqrt{3}\lambda_{so}q_{x}(1-\cos\frac{3k_{y}}{2})\tau_{0}s_{z}\sigma_{z}\nonumber\\
&&+\lambda_{\nu}\tau_{z}s_{0}\sigma_{z}-\mu\tau_{z}s_{0}\sigma_{0}\nonumber\\
&&+\frac{1+\gamma}{2}(M_{x}\tau_{z}s_{x}+M_{y}\tau_{0}s_{y})\sigma_{0}\nonumber\\
&&+\frac{1-\gamma}{2}(M_{x}\tau_{z}s_{x}+M_{y}\tau_{0}s_{y})\sigma_{z}\nonumber\\
&&-\Delta\tau_{y}s_{y}\sigma_{0}.
\end{eqnarray}
We will treat $\mathcal{H}_{1}$ as a perturbation, which is justified at least when
the parameters in $\mathcal{H}_{1}$ are all much smaller than $t$. One can see that
the two terms in $\mathcal{H}_{0}$ have a momentum dependence similar to the Su-Schrieffer-Heeger (SSH)
model~\cite{Su1979}, but with the dimension of the Hamiltonian being increased from $2$ to $8$. On
the other hand, $\mathcal{H}_{0}$ is independent of $q_{x}$. This
implies that each $y$-normal edge may harbor a zero-energy flat
band with four-fold degeneracy if periodic  boundary conditions are imposed in the
$x$ direction. To confirm
this, we focus on the upper $y$-normal edge for illustration.

To simplify the derivation, we consider a half-infinity sample with the boundary corresponding
to the upper beard edge. Accordingly, a natural basis is $\Psi_{q_{x}}=(c_{1,A,q_{x}},c_{1,B,q_{x}},c_{2,A,q_{x}},c_{2,B,q_{x}},
...,c_{n,A,q_{x}},c_{n,B,q_{x}},...)^{T}$ with $c_{n,\alpha,q_{x}}=
(c_{n,\alpha,q_{x},\uparrow},c_{n,\alpha,q_{x},\downarrow},c_{n,\alpha,-q_{x},\uparrow}^{\dag},c_{n,\alpha,-q_{x},\downarrow}^{\dag})$,
where $\alpha=\{A, B\}$.
Under this basis, the matrix form of $\mathcal{H}_{0}$ reads
\begin{eqnarray}
\mathcal{H}_{0}=\left(
        \begin{array}{cccccc}
          0 & t\tau_{z}s_{0} & 0 & 0 & 0 &  \cdots \\
          t\tau_{z}s_{0} & 0 & 2t\tau_{z}s_{0} & 0 & 0 &  \cdots \\
          0 & 2t\tau_{z}s_{0} & 0 & t\tau_{z}s_{0} & 0 &  \cdots \\
          0 & 0 & t\tau_{z}s_{0} & 0 & 2t\tau_{z}s_{0} &  \cdots \\
          0 & 0 & 0 & 2t\tau_{z}s_{0} & 0  &  \cdots \\
          \vdots & \vdots & \vdots & \vdots & \vdots &  \ddots \\
        \end{array}
      \right).\,
\end{eqnarray}
Here each ``0'' element denotes a four-by-four null matrix. The wave functions of the zero-energy bound states are determined by solving
the eigenvalue equation $\mathcal{H}_{0}|\Psi_{\alpha}\rangle=0$. By observation,
one can notice that $\tau_{z}$ and $s_{z}$ both commute with $\mathcal{H}_{0}$, so
$|\Psi_{\alpha}\rangle$ can be assigned with the form
\begin{eqnarray}
|\Psi_{\tau s}\rangle=(\psi_{1A},\psi_{1B},\psi_{2A},\psi_{2B},...)^{T}\otimes |\tau_{z}=\tau\rangle\otimes|s_{z}=s\rangle,\qquad
\end{eqnarray}
where $|\tau_{z}=\tau\rangle$ and $|s_{z}=s\rangle$ with $\tau=\pm 1$ and $s=\pm1$ correspond to the
two eigenstates of $\tau_{z}$ and $s_{z}$, respectively. Solving
the eigenvalue equation $\mathcal{H}_{0}|\Psi_{\tau s}\rangle=0$ is equivalent to solving
the following iterative equations,
\begin{eqnarray}
&&t\psi_{1B}=0,\nonumber\\
&&t\psi_{1A}+2t\psi_{2A}=0,\nonumber\\
&&2t\psi_{1B}+t\psi_{2B}=0,\nonumber\\
&&...\nonumber\\
&&t\psi_{nA}+2t\psi_{(n+1)A}=0,\nonumber\\
&&2t\psi_{nB}+t\psi_{(n+1)B}=0,\nonumber\\
&&...,
\end{eqnarray}
 According to the iterative structure, it is easy to find
\begin{eqnarray}
\psi_{(n+1)A}=-\frac{1}{2}\psi_{nA}, \quad \psi_{nB}=0.
\end{eqnarray}
Therefore, the eigenvectors take the form
\begin{eqnarray}
|\Psi_{\tau s}\rangle&=&\mathcal{N}(1,0,-\frac{1}{2},0,...,
(-\frac{1}{2})^{(n-1)},0,...)^{T}\nonumber\\
&&\otimes |\tau_{z}=\tau\rangle\otimes|s_{z}=s\rangle,
\end{eqnarray}
where $\mathcal{N}$ denotes the normalization constant. According to the normalization condition
$\langle \Psi_{\tau s} |\Psi_{\tau s}\rangle=1$,  simple calculations reveal
\begin{eqnarray}
\mathcal{N}^{2}\sum_{n=0}^{\infty}\frac{1}{2^{2n}}=\mathcal{N}^{2}\frac{1}{1-\frac{1}{4}}=\frac{4}{3}\mathcal{N}^{2}=1,
\end{eqnarray}
so $\mathcal{N}=\frac{\sqrt{3}}{2}$. As $\psi_{nA}$ decays exponentially with the increase of $n$,
the existence of four such eigenvectors indicates the existence of four zero-energy bound states, confirming
the correctness of the simple analysis based on the connection to SSH model. It is worth noting
that the TI has only one pair of gapless helical states on a given edge, so there should
exist only two degenerate zero-energy bound states at $q_{x}=0$. Here the existence of four zero-energy bound states
originates from the doubling due to the introduction of particle-hole redundancy.
The low-energy Hamiltonian on the upper $y$-normal beard edge is then obtained by
projecting $\mathcal{H}_{1}$ onto the four-dimensional subspace spanned by the four
orthogonal eigenstates. Put it explicitly, the matrix elements of the low-energy boundary
Hamiltonian are given by
\begin{eqnarray}
[\mathcal{H}_{\rm BdG,b}(q_{x})]_{\tau s,\tau's'}=\langle \Psi_{\tau s}|\mathcal{H}_{1}(q_{x})|\Psi_{\tau' s'}\rangle.
\end{eqnarray}
It is worth noting that here $\mathcal{H}_{1}(q_{x})$ is also an infinitely large matrix, and its form
is given by a partial Fourier transform of $\mathcal{H}_{1}(q_{x},k_{y})$ in the $y$ direction. By some straightforward calculations
and choosing $(|\Psi_{11}\rangle,|\Psi_{1-1}\rangle,|\Psi_{-11}\rangle,|\Psi_{-1-1}\rangle)^{T}$
as the basis for the low-energy boundary Hamiltonian,
one can obtain (more details on how to determine each term in the low-energy Hamiltonian
can be found in Ref.~\cite{Zhu2021sublattice})
\begin{eqnarray}
\mathcal{H}_{\rm BdG,b}(q_{x})&=&vq_{x} \tau_{0}s_z
   +M_{x} \tau_{z} s_{x}+M_y \tau_{0}s_{y}\nonumber\\
   &&-(\mu - \lambda_{\nu}) \tau_{z}s_{0}-\Delta\tau_{y}s_{y},
\end{eqnarray}
where $v=3\sqrt{3}\lambda_{\rm so}$. Without the superconductivity (the particle-hole redundancy
is accordingly removed)
and staggered potential,  the low-energy boundary Hamiltonian reduces
to the form in Eq.(\ref{lowbeard}).
The boundary energy spectra associated with this boundary Hamiltonian
read
\begin{eqnarray}
E(q_{x})=\pm\sqrt{F\pm2\sqrt{G}},
\end{eqnarray}
where $F=v^{2}q_{x}^{2}+M^{2}+(\mu-\lambda_{\nu})^{2}+\Delta^{2}$
and $G=(\mu-\lambda_{\nu})^{2}(v^{2}q_{x}^{2}+M^{2})+M^{2}\Delta^{2}$.
The gap of the boundary energy spectra gets closed at $q_{x}=0$ when
the following condition is fulfilled,
\begin{eqnarray}
M=\sqrt{(\mu-\lambda_{\nu})^{2}+\Delta^{2}}.
\end{eqnarray}

\section{Low-energy boundary Hamiltonian on the zigzag edge}

Following the same spirit, we can derive the low-energy boundary Hamiltonian
on the upper $y$-normal zigzag edge. Since numerical results show that the boundary
Dirac point on the $y$-normal zigzag edge is located at $k_{x}=\pi/\sqrt{3}$,
we similarly expand the Hamiltonian up to the linear order
in momentum, which then gives
\begin{eqnarray}
\mathcal{H}_{\rm BdG}(q_{x}',k_{y})&=&t(-\sqrt{3}q_{x}'\cos\frac{k_{y}}{2}+\cos k_{y})\tau_{z}s_{0}\sigma_{x}\nonumber\\
&&+t(\sqrt{3}q_{x}'\sin\frac{k_{y}}{2}+\sin k_{y})\tau_{z}s_{0}\sigma_{y}\nonumber\\
&&+2\lambda_{\rm so}(-\sqrt{3}q_{x}'-2\cos\frac{3k_{y}}{2})\tau_{0}s_{z}\sigma_{z}\nonumber\\
&&+\lambda_{\nu}\tau_{z}s_{0}\sigma_{z}-\mu\tau_{z}s_{0}\sigma_{0}\nonumber\\
&&+\frac{1+\gamma}{2}(M_{x}\tau_{z}s_{x}+M_{y}\tau_{0}s_{y})\sigma_{0}\nonumber\\
&&+\frac{1-\gamma}{2}(M_{x}\tau_{z}s_{x}+M_{y}\tau_{0}s_{y})\sigma_{z}\nonumber\\
&&-\Delta\tau_{y}s_{y}\sigma_{0},
\end{eqnarray}
where $q_{x}'$ denotes a small momentum measured from $k_{x}=\pi/\sqrt{3}$.
Similarly, we decompose the Hamiltonian into two parts,
$\mathcal{H}=\mathcal{H}_{0}+\mathcal{H}_{1}$, with
\begin{eqnarray}
\mathcal{H}_{0}(q_{x}',k_{y})&=&t\cos k_{y}\tau_{z}s_{0}\sigma_{x}+t\sin k_{y}\tau_{z}s_{0}\sigma_{y}\nonumber\\
&&-4\lambda_{\rm so}\cos\frac{3k_{y}}{2}\tau_{0}s_{z}\sigma_{z},\nonumber\\
\mathcal{H}_{1}(q_{x}',k_{y})&=&-\sqrt{3}tq_{x}'\cos\frac{k_{y}}{2}\tau_{z}s_{0}\sigma_{x}
+\sqrt{3}tq_{x}'\sin\frac{k_{y}}{2}\tau_{z}s_{0}\sigma_{y}\nonumber\\
&&-2\sqrt{3}\lambda_{\rm so}q_{x}'\tau_{0}s_{z}\sigma_{z}\nonumber\\
&&+\lambda_{\nu}\tau_{z}s_{0}\sigma_{z}-\mu\tau_{z}s_{0}\sigma_{0}\nonumber\\
&&+\frac{1+\gamma}{2}(M_{x}\tau_{z}s_{x}+M_{y}\tau_{0}s_{y})\sigma_{0}\nonumber\\
&&+\frac{1-\gamma}{2}(M_{x}\tau_{z}s_{x}+M_{y}\tau_{0}s_{y})\sigma_{z}\nonumber\\
&&-\Delta\tau_{y}s_{y}\sigma_{0}.
\end{eqnarray}
Also focusing on the upper $y$-normal boundary, the change from a beard edge to a zigzag edge
is accompanied with  the change of terminating sublattice from sublattice A to sublattice B. For simplicity,
we also consider the half-infinity geometry, and then the corresponding basis becomes
$\Psi_{q_{x}'}=(c_{1,B,q_{x}'},c_{2,A,q_{x}'},c_{2,B,q_{x}'},c_{3,A,q_{x}'},c_{3,B,q_{x}'},
...)^{T}$. Under this basis, the corresponding matrix form of $\mathcal{H}_{0}$ reads
\begin{eqnarray}
\mathcal{H}_{0}=\left(
        \begin{array}{cccccc}
          0 & T_{1} & 0 & 0 & 0 &  \cdots \\
          T_{1}^{\dag} & 0 & T_{1} & 0 & 0 &  \cdots \\
          0 & T_{1}^{\dag} & 0 & T_{1} & 0 &  \cdots \\
          0 & 0 & T_{1}^{\dag} & 0 & T_{1} &  \cdots \\
          0 & 0 & 0 & T_{1}^{\dag} & 0  &  \cdots \\
          \vdots & \vdots & \vdots & \vdots & \vdots &  \ddots \\
        \end{array}
      \right),\,
\end{eqnarray}
where now each ``0'' element in $\mathcal{H}_{0}$ is an eight-by-eight null matrix,
and
\begin{eqnarray}
T_{1}&=&\left(
        \begin{array}{cc}
          2\lambda_{\rm so}\tau_{0}s_{z} & 0 \\
          t\tau_{z}s_{0} & -2\lambda_{\rm so}\tau_{0}s_{z} \\
        \end{array}
      \right).
\end{eqnarray}
As here $\tau_{z}$ and $s_{z}$ also commute with $\mathcal{H}_{0}$,  the wave functions
for zero-energy bound states can also be assigned with the from
\begin{eqnarray}
|\Psi_{\tau s}\rangle&=&(\psi_{1B},\psi_{2A},\psi_{2B},\psi_{3A},\psi_{3B},...)^{T}\nonumber\\
&&\otimes |\tau_{z}=\tau\rangle\otimes|s_{z}=s\rangle.
\end{eqnarray}
The eigenvalue equation $\mathcal{H}_{0}|\Psi_{\tau s}\rangle=0$ leads to the following
iterative equations,
\begin{eqnarray}
&&2\lambda_{\rm so,s}\psi_{2B}=0,\nonumber\\
&&t_{\tau }\psi_{2B}-2\lambda_{\rm so, s}\psi_{3A}=0,\nonumber\\
&&2\lambda_{\rm so,s}\psi_{1B}+t_{\tau }\psi_{2A}+2\lambda_{\rm so, s}\psi_{3B}=0,\nonumber\\
&&-2\lambda_{\rm so, s}\psi_{2A}+t_{\tau }\psi_{3B}-2\lambda_{\rm so, s}\psi_{4A}=0,\nonumber\\
&&...\nonumber\\
&&2\lambda_{\rm so,s}\psi_{(n-1)B}+t_{\tau }\psi_{nA}+2\lambda_{\rm so,s}\psi_{(n+1)B}=0,\nonumber\\
&&-2\lambda_{\rm so, s}\psi_{nA}+t_{\tau}\psi_{(n+1)B}-2\lambda_{\rm so,s}\psi_{(n+2)A}=0,\nonumber\\
&&...,
\end{eqnarray}
where $t_{\tau}=t\tau$ and $\lambda_{\rm so,s}=\lambda_{\rm so}s$. The solutions are found to take
the form (more details can be found in Ref.~\cite{Zhu2021sublattice})
\begin{eqnarray}
|\Psi_{\tau s}\rangle&=&\mathcal{N}(\eta_{\tau s},1,0,0,\xi\eta_{\tau s},\xi,0,0,\xi^{2}\eta_{\tau s},\xi^{2},...)^{T}\nonumber\\
&&\otimes|\tau_{z}=\tau\rangle\otimes|s_{z}=s\rangle,
\end{eqnarray}
where
\begin{eqnarray}
\xi&=&\frac{\sqrt{t^{2}(t^{2}+16\lambda_{\rm so}^{2})}-(t^{2}+8\lambda_{\rm so}^{2})}{8\lambda_{\rm so}^{2}},\nonumber\\
\mathcal{N}&=&\sqrt{\frac{{1-\xi^{2}}}{1+\eta^{2}}},\nonumber\\
\eta&=&\frac{4t\lambda_{\rm so}}{\sqrt{t^{2}(t^{2}+16\lambda_{\rm so}^{2})}-t^{2}},
\end{eqnarray}
and $\eta_{\tau s}=-\tau s\eta$. Similarly, projecting $\mathcal{H}_{1}$ onto the four-dimensional
subspace spanned by the four orthogonal wave functions associated with the
four zero-energy bound states, one can obtain the low-energy boundary Hamiltonian,
which reads
\begin{eqnarray}
\mathcal{H}_{\rm BdG;z}(q_{x}')& = & v'q_{x}' \tau_{0}s_z
   +\frac{1-\gamma\eta^{2}}{1+\eta^{2}}(M_{x} \tau_{z} s_{x}+M_y \tau_{0}s_{y})\nonumber\\
   &&-(\mu +\frac{\eta^{2}-1}{1+\eta^{2}}\lambda_{\nu}) \tau_{z}s_{0}-\Delta\tau_{y}s_{y},\label{full}
\end{eqnarray}
where
\begin{eqnarray}
v'=\frac{2\sqrt{3}t\eta+2\sqrt{3}\lambda_{so}(\eta^{2}-1)}{1+\eta^{2}}.
\end{eqnarray}
For real materials, it is common that $t\gg \lambda_{\rm so}$. When
$t\gg \lambda_{\rm so}$, one finds
\begin{eqnarray}
\eta&=&\frac{4t\lambda_{\rm so}}{\sqrt{t^{2}(t^{2}+16\lambda_{\rm so}^{2})}-t^{2}}\nonumber\\
&\approx&\frac{4t\lambda_{\rm so}}{(t^{2}+8\lambda_{\rm so}^{2})-t^{2}}\nonumber\\
&=&\frac{t}{2\lambda_{\rm so}}\gg1.
\end{eqnarray}
In this limit, $v'\approx2v$, and the boundary Hamiltonian for the zigzag edge can be simplified as
\begin{eqnarray}
\mathcal{H}_{\rm BdG;z}(q_{x}')& \approx & v'q_{x}' \tau_{0}s_z
   -\gamma(M_{x} \tau_{z} s_{x}+M_y \tau_{0}s_{y})\nonumber\\
   &&-(\mu +\lambda_{\nu}) \tau_{z}s_{0}-\Delta\tau_{y}s_{y}.
\end{eqnarray}
The corresponding boundary energy spectra read
\begin{eqnarray}
E(q_{x}')=\pm\sqrt{F'\pm2\sqrt{G'}},
\end{eqnarray}
where $F'=v'^{2}q_{x}^{2}+\gamma^{2}M^{2}+(\mu+\lambda_{\nu})^{2}+\Delta^{2}$
and $G'=(\mu+\lambda_{\nu})^{2}(v'^{2}q_{x}^{2}+\gamma^{2}M^{2})+\gamma^{2}M^{2}\Delta^{2}$.
The gap of the boundary energy spectra gets closed at $q_{x}'=0$ when
the following condition is fulfilled,
\begin{eqnarray}
|\gamma| M=\sqrt{(\mu+\lambda_{\nu})^{2}+\Delta^{2}}.\label{criterion}
\end{eqnarray}
If one determines the gap-closing condition according to the Hamiltonian (\ref{full}),
one only needs to do the replacement, $\gamma\rightarrow(1-\gamma\eta^{2})/(1+\eta^{2})$,
and $\lambda_{\nu}\rightarrow\lambda_{\nu}(\eta^{2}-1)/(1+\eta^{2})$, namely, the criterion
(\ref{criterion}) for gap closure is simply modified as
\begin{eqnarray}
|\frac{1-\gamma\eta^{2}}{1+\eta^{2}}| M=\sqrt{(\mu+\frac{\eta^{2}-1}{1+\eta^{2}}\lambda_{\nu})^{2}+\Delta^{2}}.\label{accurate}
\end{eqnarray}

\bibliography{dirac}

\end{document}